\documentclass[12pt, draftclsnofoot, onecolumn]{IEEEtran}
\usepackage{cite}
\usepackage{graphicx}
\usepackage{array}
\usepackage{tikz}
\usepackage{tikz-3dplot}
\usetikzlibrary{shapes.geometric}
\usetikzlibrary{decorations.pathreplacing}
\usetikzlibrary{positioning}
\usetikzlibrary{matrix}
\usepackage{pgfplots}
\usepackage{pstool}  
\usepackage{tikz}
\usetikzlibrary{automata,positioning}

\usepackage[linesnumbered,algoruled,boxed,lined]{algorithm2e}
\usepackage[absolute]{textpos}
\usepackage{threeparttable}
\usepackage[strict]{changepage}
\usepackage{bm,upgreek} 
\usepackage{amssymb}
\usepackage{comment}
\usepackage{bbm}
\usepackage{hhline}
\usepackage{amsthm}

\usepackage{lipsum} 
\usepackage{enumitem}
\usepackage{soul}

\renewcommand{\thefootnote}{\fnsymbol{footnote}}
\ifCLASSINFOpdf
\else
\fi
\usepackage{amsmath}
\ifCLASSOPTIONcompsoc
  \usepackage[caption=false,font=normalsize,labelfont=sf,textfont=sf]{subfig}
\else
  \usepackage[caption=false,font=footnotesize]{subfig}
\fi
\newcommand{\given}{\,\vert\,}
\newcommand{\Given}{\,\bigg\vert\,}

\pgfplotsset{compat=1.14}

\begin{document}
%
\title{A Framework for Automated Cellular Network Tuning with Reinforcement Learning}

%

\author{Faris~B.~Mismar,~\IEEEmembership{Senior Member,~IEEE,}
	 Jinseok~Choi,~\IEEEmembership{Student Member,~IEEE,}
        and~Brian~L.~Evans,~\IEEEmembership{Fellow,~IEEE}
\thanks{The authors are with the Wireless Networking and Communications Group, Department of Electrical and Computer Engineering, The University of Texas at Austin, Austin, TX, 78712, USA e-mail: \{faris.mismar, jinseokchoi89\}@utexas.edu and bevans@ece.utexas.edu.} %
\thanks{This paper is an expanded journal version of \cite{faris_volte} and \cite{faris_fm}.}
}

\maketitle

\begin{abstract}

Tuning cellular network performance against always occurring wireless impairments can dramatically improve reliability to end users. In this paper, we formulate cellular network performance tuning as a reinforcement learning (RL) problem and provide a solution to improve the performance for indoor and outdoor environments. By leveraging the ability of $Q$-learning to estimate future performance improvement rewards, we propose two algorithms: (1) {\color{black} closed loop power control (PC) for downlink voice over LTE (VoLTE)} and (2) self-organizing network (SON) fault management. The VoLTE PC algorithm uses RL to adjust the indoor base station transmit power so that the signal to interference plus noise ratio (SINR) of a user equipment (UE) meets the target SINR. {\color{black}It does so without the UE having to send power control requests.}  The SON fault management algorithm uses RL to improve the performance of an outdoor base station cluster by resolving faults in the network through configuration management. Both algorithms exploit measurements from the connected users, wireless impairments, and relevant configuration parameters to solve a non-convex performance optimization problem using RL. Simulation results show that our proposed RL based algorithms outperform the industry standards today in realistic cellular communication environments. 


\end{abstract}


%
\IEEEpeerreviewmaketitle

\section{Introduction} 
%
%
%
%

The tuning of network performance aims at providing the end user with a \textit{quality of experience} (QoE) suitable for the desired service.  With a projection of over 2.8 billion smartphone users globally by 2020 \cite{statista1}, 
demand patterns are likely to continue to change.  Demands have shifted towards reliable packetized voice and applications with higher data rates and lower latencies \cite{statista2}.  This shift has created a need to proactively tune cellular networks for performance while minimizing fault resolution times.  
In this paper, we propose a framework to automatically tune a cellular network through the use of reinforcement learning (RL).



 

While cellular data applications are made resilient against wireless impairments such as blockage, interference, or failures in network elements by means of retransmissions and robust modulation and coding, delay-sensitive applications such as voice or low latency data transfer may not always benefit from retransmission since it increases delays and risk of duplication.  These applications need to become resilient through other means.  {\color{black} Further, network operational faults (such as changes in antenna azimuth or tilts) may impact the cellular coverage.  Such impacts on coverage may deteriorate the QoE for users requiring these delay-sensitive applications.}

We devise RL-based algorithms to improve downlink performance in practical cellular environments: indoor voice-over-LTE \textit{power control} (PC) and outdoor automated fault management.  The technology of focus is the fourth generation wireless communications or \textit{long term evolution} (4G LTE) or fifth generation wireless communications (5G).

\subsection{Related Work and Motivation}

An improved decentralized $Q$\nobreakdash-learning algorithm to reduce interference in LTE femtocells was derived in \cite{5983301} with a comparison against various PC algorithms including open loop PC. The Kullback-Leibler divergence and entropy constraints in deep RL was introduced in \cite{infoml}.  
The first deep $Q$-learning framework was successfully introduced to learn control policies directly using RL in \cite{mnih2013playing}. The framework outperformed human experts in three out of seven trials.  It required a low-dimensional action space so that the reinforcement learning agent could enumerate all possible actions at the current state and perform the inference.   


Focusing on throughput analysis, $Q$\nobreakdash-learning based PC for indoor LTE femtocells with an outdoor macro cell was performed in \cite{7442078}.  The \textit{user equipment} (UE) reported its  \textit{signal to interference plus noise ratio} (SINR), which was used as a performance measure, to the serving cell.  A central controller was introduced to resolve the issue of communicating base stations.  Two assumptions were made: (1) the downlink PC was achieved over shared data channels and (2) the scheduler was aware ahead of time about the channel condition for the upcoming user to perform PC. In this paper, we do not make these assumptions to keep our setup realistic.


Deep learning in mobile and wireless networking with interference alignment was studied in \cite{HeZYZYLZ17}. As relaxed \textit{channel state information} (CSI) assumptions were made prior to this study (e.g., block-fading channels or time-invariant channels), a time-varying channel was proposed. An assumption, however, was made that the CSI transition matrix was identical across all users, which we overcome in a multi-cell environment. 
In addition, the two-dimensional convolutional neural network used in simulations \cite{HeZYZYLZ17} invites the creation of unfounded spatial relationships between learning features (also known as \textit{local connection patterns} \cite{726791}), which we avoid in our design of our deep neural network.

A means to improve the handover execution success rate using supervised machine learning was devised  in \cite{faris}.  This approach, however, did not use RL, which has the ability to learn from previous actions.  It instead depended on coherence time for the validity of the approach. A method for extracting the knowledge base from solved fault troubleshooting cases was proposed in \cite{KHATIB20157549}. It used data mining and supervised learning techniques, fuzzy logic, and expert opinions to define performance measurements and targets. On the other hand, we use reinforcement learning to derive a near-optimal policy to map actions to be taken by the self-healing functionality in response to select common number of faults in the network. 

Downlink closed loop PC was last implemented in 3G \textit{universal mobile telecommunications system} (UMTS) \cite{3gpp25214}.  It rapidly adjusted the transmit power of a radio link of a dedicated traffic channel to match the target SINR.  This technique is not present in 4G LTE or 5G due to the absence of dedicated traffic channels for packet data sessions.  However, the introduction of \textit{semi-persistent scheduling} (SPS) in 4G LTE has created a virtual sense of a dedicated downlink traffic channel for \textit{voice over LTE} (VoLTE) on which closed-loop PC can be performed.  This scheduling is at least for the length of one voice frame---which is on order of tens of LTE \textit{transmit time intervals} (TTIs).
In \cite{5638376}, the authors proposed uplink closed loop PC implementation for LTE and used fractional path loss compensation to improve the system performance.  There was no reference to machine learning or RL in general, where obtaining pertinent training data for the machine learning models may be a challenge.

\mbox{$Q$-learning} as part of the SON implementation for mobile load balancing and mobility optimization for cell reselection and handovers in single-transceiver cells was devised in \cite{7393587}. We, on the other hand, introduce multiple transceiver cells, or \textit{multiple-input multiple-output} (MIMO), which is a fundamental setup for present and futuristic network deployments. 

Deep RL learning was studied in \cite{8303773} in a dynamic multichannel access with an objective to find a policy that maximizes the expected long-term number of successful transmissions.  Near-optimal performance was achieved using deep RL without knowing the system statistics.  The use of RL in \textit{device-to-device communications} (D2D) was studied in \cite{7517324,8410619}.  In \cite{7517324}, an autonomous operation of D2D pairs in a heterogeneous cellular network was studied  where a multi-agent $Q$-learning algorithm was developed where each device becomes a learning agent whose task is to learn its best policy.  An attempt to improve spectral efficiency in D2D communications in cloud radio access networks using RL was made in \cite{8410619}.

Unlike prior work, our proposed closed loop PC addresses voice instead of data bearers, exploits the existence of SPS in 4G LTE, and uses RL to achieve the objective from within the base station.  It does so without the need of explicit commands from the UEs.  Our proposed SON fault management employs automation through RL instead of through a series of explicit policies, workflows, and SON functions, which are the case in SON today \cite{6666642}. RL is well-suited to problems which include a long-term versus short-term reward trade-off \cite{Sutton}.  This includes cellular network tuning.

\begin{figure*}[!t]
\centering
\begin{tikzpicture}[style=thick,scale=0.5]
\matrix[matrix of nodes, nodes={draw, rectangle, text width= 12em,
text height=1.5ex, text depth=0.25ex,
align = center}, row sep = 1ex,column sep = -6em] 
(mx2){
        & 
          \node[rectangle, draw, rounded corners, text width=8em, text centered, minimum height=1em, anchor=west](a0){\small Indoor}; 
          \node[rectangle, draw, rounded corners, text width=8em, text centered, minimum height=1em, right=of a0](a1){\small Outdoor};         
          \node[rectangle, draw, rounded corners, text width=8em, text centered, minimum height=1em, right=of a1](a2){\small Hybrid};         
        \\
        &
            \node[rectangle, draw, rounded corners, text width=8em, text centered, minimum height=1em, below=of a0] (b0) {\small Power Control};
            \node[rectangle, draw, rounded corners, text width=8em, text centered, minimum height=1em] (b1) [right=of b0] {\small Fault Handling};
            \node[rectangle, draw, rounded corners, text width=8em, text centered, minimum height=1em] (b2) [right=of b1]  {\small Etc.};
            
        \\
        &
            \node[rectangle, draw, rounded corners, text width=8em, text centered, minimum height=1em, right=of b0.south] (c0) {\small Tabular RL};
            \node[rectangle, draw, rounded corners, text width=8em, text centered, minimum height=1em, right=of c0] (c1) {\small Deep RL};
            
        \\
        &
            \node[rectangle, draw, rounded corners, text width=10em, fill=gray!40, text centered, minimum height=1em, below=of a1] (d0) {\small Near-optimal actions};            
        \\
    };
    
    \foreach \x in {0,1,2}
        \foreach \y in {0,1,2}
          \draw[-latex, line width=1pt] (a\x.270)--(b\y.90);
    \linespread{1}
    \foreach \y in {0,1,2}
        \foreach \z in {0,1}
            \draw[-latex, line width=1pt] (b\y.270)-- node[text width=7em, rotate=325, align=center] {\ifthenelse{\z=0 \AND \y=0}{\smaller  actions  \\ observations}{}} (c\z.90);


\draw[draw=black] (-17,5) rectangle ++(29,3) node[right,above,xshift=-1.6cm] {Radio Environment};
\draw[draw=black] (-17,1) rectangle ++(29,3) node[right,above,xshift=-1.1cm] {Problem Set};
\draw[draw=black] (-17,-3.75) rectangle ++(29,3) node[right,above,xshift=-0.5cm] {Agent};
    
\end{tikzpicture}%
\vspace*{-1em}
\caption{{\color{black}Generalized framework diagram.  To solve RRM-based problems, traverse the diagram from the top downwards.}}\label{fig:framework}
\end{figure*}
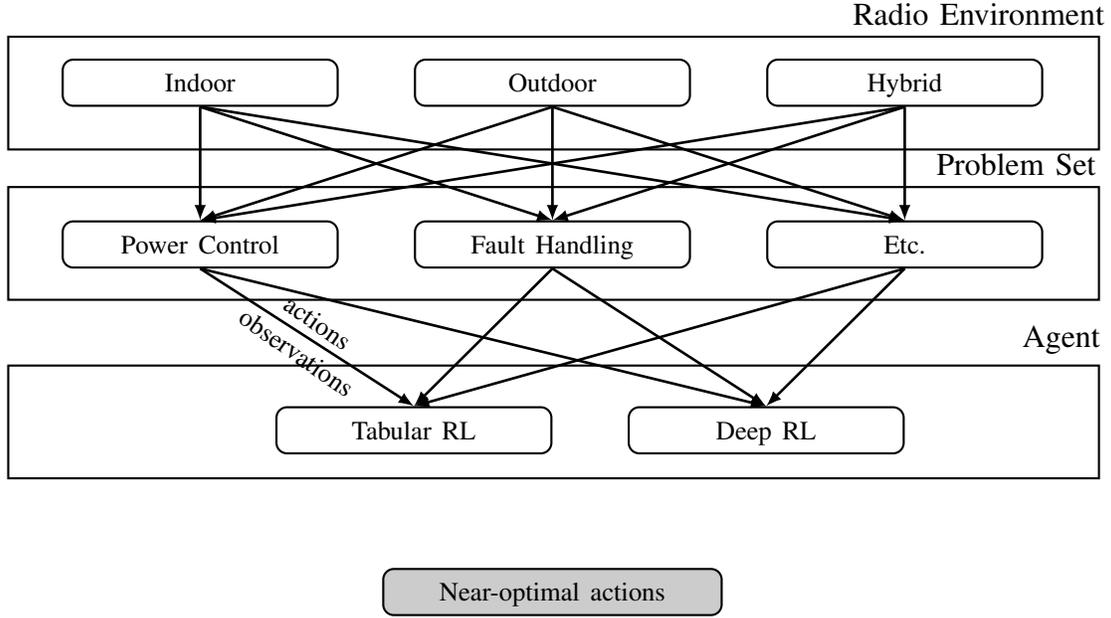

{\color{black}
Faults in cellular networks cause degraded service and can lead to system failure.  These {\color{black} degradations cause} poor end-user QoE.  An intelligent fault management algorithm that can handle faults as they occur becomes a necessity to improve end-user QoE.  Furthermore, cellular network tuning is a major component of the network operating expenditure \cite{1427028}. Although we solve the indoor and outdoor problems by using different types of RL, the main idea here is similar for the two problems: learn a near-optimal recovery policy in the absence of sufficient training data.

Tuning the cellular network radio parameters is commonly known as \textit{radio resource management} (RRM).  {\color{black} A generalized system diagram of our proposed framework which we use in RRM is in Fig.~\ref{fig:framework}.  In this generalized diagram, we show various cellular network environments, a set of RRM problems that can be resolved, and the choice of RL to solve these problems.  To solve such a problem, traverse the diagram from the top.}  The use of RL to perform real-time RRM is therefore valuable in maintaining the end-user QoE against impairments.  
While the aforementioned indoor and outdoor problems are RRM problems that can be solved with RL {\color{black} as shown in Fig.~\ref{fig:framework}}, there are other reasons why we choose these two problems:
\begin{enumerate}
    \item The two problems are well-defined.
    \item The formulation of both problems yields solutions that are standard-compliant \cite{3gpp36213,itum3400}.
    \item Both problems are about optimizing an objective to enhance the end-user QoE particularly against wireless signal impairments and network operational faults.
\end{enumerate}
Therefore, the approaches we use to solve the two problems could be applied to a wide variety of cellular network tuning problems.
}

\subsection{Contributions} 

We use RL to solve a non-convex constrained SINR optimization problem in our investigation of VoLTE PC and network fault management as part of SON.  The  motivation of using RL for such problems to create a framework is its ability to formulate a policy that can improve the performance of the serving base stations.  {\color{black} The policy describes the behavior of a decision-based agent \color{black} which is the base station in the VoLTE PC problem.}  VoLTE PC using RL functions constitutes a closed loop PC which ensures that the serving base station radio link power is constantly tuned so that the target downlink SINR is met.  {\color{black} We use the UE measurement reports  of their received downlink SINR, which are sent to the base station, and the indoor network topology to develop the algorithm. }

{\color{black} We also propose using RL on SON fault management to autonomously and intelligently resolve the impact of impairments on downlink throughput as experienced by UEs.  In SON fault management, the decision-based agent is the performance technician, who looks after the end-user QoE.} 

To demonstrate the performance of the proposed RL-based algorithms, we adopt a realistic simulation environment.  Simulation results show that our RL-based algorithms improve the performance of the cellular network as measured by standard practice performance measures.  With the introduction of edge computing to current 4G and 5G cellular networks \cite{etsiedge}, the proposed algorithms can readily be deployed in these communication networks. 

Our main contributions are summarized as follows:
\begin{itemize}
    \item We adopt RL to solve performance tuning problems in a scalable cellular network beyond the \textit{physical layer} (PHY).
    \item We demonstrate that the problem formulation applies to both indoor and outdoor environments alike.
    \item We show that our derived lower bound loss in effective SINR is sufficient for power control purposes in practical cellular networks.
    \item We propose RL-based performance tuning framework that operates on upper layer protocols and outperform current industry standards.
\end{itemize}

The remainder of this paper is organized as follows. In Section~\ref{sec:network}, we discuss the cellular network and describe the network model and the signal model in detail.  In Section~\ref{sec:reinf}, we discuss reinforcement learning and its usage in our framework for cellular network tuning.  In Section~\ref{sec:algorithms}, we propose RL-based algorithms along with a few industry standard and baseline algorithms for comparison.  We show our performance measures in Section~\ref{sec:measures}, which are used to quantitatively benchmark the various algorithms.  In Section~\ref{sec:results}, we show the results of our proposed algorithms based on the selected performance measures.  We conclude the paper in Section~\ref{sec:conclusion}.

\indent \textit{Notation:} Boldface lower and upper case symbols represent column vectors and matrices, respectively. Calligraphic letters are for sets.  The cardinality of a set is $\vert\cdot\vert$.  The expectation operator is $\mathbb{E}[\cdot]$.  The $\triangleq$ symbol means equal by definition. The indicator function $\mathbbm{1}_{(\cdot)}$ is equal to one if the condition in the parentheses is true and zero if false.  We use the notation $\mathbb{F}_2^M$ to represent an $M$-dimensional vector in a binary finite field.  Finally, an $M$-by-$N$ matrix whose elements are real numbers is $\mathbb{R}^{M\times N}$.  

\section{Cellular Network}\label{sec:network}

\subsection{Network Model}

\begin{figure}[!t]
\centering
\subfloat[Indoor]{
\begin{tikzpicture}[scale=0.7,font=\scriptsize, triangle/.style = {fill=black!20, regular polygon, regular polygon sides=3}]
\draw [thin, black, latex-latex] (-2,0) -- (2,0)      
        node [right, black] {$x$ pos (m)};              

    \draw [thin, black, latex-latex] (0,-2) -- (0,2)      
        node [right, black] {$y$ pos (m)};              
\draw[help lines, color=gray!80, dashed] (-1.5,-1.5) grid (1.5,1.5);
\node[triangle,fill,inner sep=1pt,color=red](a) at (0,0) {};
\node[circle,fill,inner sep=1pt,color=black](x) at (-1,0) {};
\node[circle,fill,inner sep=1pt,color=black](y) at (1,0) {};
\node[circle,fill,inner sep=1pt,color=black](z) at (0,-1) {};
\node[circle,fill,inner sep=1pt,color=black](t) at (0,1) {};

  \foreach \linker / \regter in {%
    {(-0.5,-0.5)}/{(0.5,0.5)}, %
    {(-0.5,0.5)}/{(0.5,1.5)}, %
     {(-0.5,0.5)}/{(0.5,-1.5)}, %
     {(-0.5,0.5)}/{(-1.5,-0.5)}, %
    {(0.5,0.5)}/{(1.5,-0.5)}%
  } { \draw[thick] \linker rectangle \regter; }
 
\end{tikzpicture}%
\label{fig:first_case}
}
\hfil
\subfloat[Outdoor]{  
\begin{tikzpicture} [scale=1.25,hexa/.style= {shape=regular polygon,regular polygon sides=6,minimum size=1.25cm, thick,draw,inner sep=0,anchor=south,rotate=30},triangle/.style = {fill=black!20, regular polygon, regular polygon sides=3}]
  \node[hexa] (h0;0) at ({(0)*sin(60)},{0*0.75}) {};
  \node[hexa] (h0;1) at ({(1/2)*sin(60)},{1*0.75}) {};
  \node[hexa] (h0;2) at ({(3/2)*sin(60)},{1*0.75}) {};
  \node[hexa] (h1;0) at ({(1)*sin(60)},{0*0.75}) {};  
  \node[hexa] (h1;1) at ({(1/2)*sin(60)},{-1*0.75}) {};
  \node[hexa] (h1;2) at ({(3/2)*sin(60)},{-1*0.75}) {};
  \node[hexa] (h2;0) at ({(2)*sin(60)},{0*0.75}) {};

\node[circle,fill,inner sep=1pt,color=black](a) at (h0;0) {};
\node[circle,fill,inner sep=1pt,color=black](b) at (h0;1) {};
\node[triangle,fill,inner sep=1pt,color=red](c) at (h1;0) {}; 
\node[circle,fill,inner sep=1pt,color=black](d) at (h0;2) {};
\node[circle,fill,inner sep=1pt,color=black](e) at (h2;0) {};
\node[circle,fill,inner sep=1pt,color=black](f) at (h1;2) {};
\node[circle,fill,inner sep=1pt,color=black](g) at (h1;1) {};

\end{tikzpicture}

  
  

%
\label{fig:second_case}
}
\caption{Cellular network layouts.  The red triangle represents the serving base station.  The black points are the neighboring base station.}
\label{fig:layout}
\end{figure}
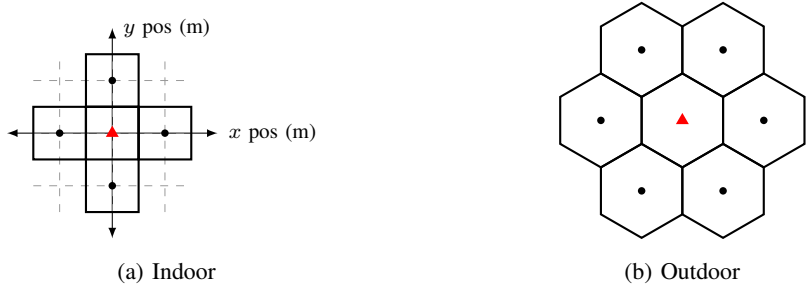

We consider an \textit{orthogonal frequency division multiplexing} (OFDM) multi-access downlink cellular network consisting of indoor and outdoor deployments.  The distribution of the indoor cells is deterministic (i.e., one base station at the center of each room in a floor plan of several adjacent rooms) as depicted in Fig.~\ref{fig:first_case}, while the distribution of the outdoor base stations is equi-distant in hexagonal structures \cite{VLS-2016} as shown in Fig.~\ref{fig:second_case}.
The users in the indoor environment are engaged in packetized voice sessions over 4G LTE, known as VoLTE, while the users in the outdoor environment are engaged in 4G LTE high speed data access.  {\color{black} We focus on packetized voice indoors since more than 60\% of the indoor high speed data access traffic comes from Wi-Fi \cite{gsma}.}

The successful reception of a VoLTE frame in the indoor environment depends on the downlink SINR as received by the UE.  We consider that if it is larger than a target SINR threshold $\gamma_\text{DL, target}$, the frame is successfully received by the UE.  
In the outdoor environment, we use the number of unresolved network faults as a proxy to the successful reception of the UEs.  The behavior of the data throughput received by each UE is governed by the industry standards of LTE \cite{3gpp36321}.

Regardless of whether indoors or outdoors, the network can either operate normally or undergo a few faults.  These faults, which can worsen the performance of the wireless signal, depend on the environment (i.e., indoor vs. outdoor).  We denote a set of these faults by $\mathcal{N}\triangleq\{\nu_i\}_{i=1}^{\vert\mathcal{N}\vert}$.  Each one of these faults can happen in the network at a finite rate $p_{\nu,i}\in[0,1]$. We study the impact of these faults on the downlink SINR.  These faults are tracked in a fault register.  With every frame having finite transmission duration, we assume that RL-based algorithms can select an action to tune the performance of the network after each frame.

\subsection{Signal Model}

{\color{black} In this model, our transmitter is the base station, while the receivers are the served UEs.} We start with our forward link budget, which at any time $t$ for the $i$-th receiving UE is written in dBm as
\begin{equation}
    P_{\text{UE}}^{(i)}[t] =  P_{\text{TX}}^{(i)}[t] + G_\text{TX} - L_\text{m} - L_{\text{p}}^{(i)}[t]  + G_\text{UE}
    \label{eq:linkbudget}
\end{equation}
where $P_{\text{UE}}^{(i)}$ is the power received by the $i$-th UE for the service-dependent allocated  \textit{physical resource blocks} (PRB) transmitted  at time $t$ by the base station at power $P_\text{TX}$, $G_\text{TX}$ is the antenna gain of the transmitter, $L_\text{m}$ is a miscellaneous loss (e.g., feeder loss and return loss), $L_{\text{p}}^{(i)}[t]$ is the time-dependent $i$-th UE path loss over the air interface for \textit{line of sight} (LOS) indoor propagation, and $G_\text{UE}$ is the UE  receiving antenna gain. 
{\color{black} For indoor settings, we use LOS propagation aligned with \cite{8330328} for sub-6 GHz transmissions.
}

Now, we compute the received SINR for the $i$-th UE at TTI $t$, $\gamma_{\text{DL}}^{(i)}[t]$, for $i \in \{1, 2, \ldots, N_\text{UE}\}$ as follows:
\begin{align}\label{eq:sinr_final}
    \gamma_{\text{DL}}^{(i)} \triangleq \frac{P_{\text{UE}}^{(i)}}{N_0 + \underbrace{\sum_{j: \mathbf{o}_j \in\mathcal{C} \setminus \{{\mathbf{o}_0}\}} {  P_{\text{UE}, \mathbf{o}_j \rightarrow i}}}_{\text{ICI}}}.
\end{align}
Here, we dropped the time index for ease of notation.  $N_0$ is the white Gaussian noise variance, $P_{\text{UE}}^{(i)}$ is defined as in (\ref{eq:linkbudget}), $\mathcal{C}$ is a set of all the base stations in the cluster, $\mathbf{o}_j$ is the coordinates of the $j$-th base station $(j \in \{1,2,\ldots,\vert\mathcal{C}\vert - 1\})$. {\color{black}Without loss of generality, we assume that $\mathbf{o}_0$ is the serving base station placed at the origin. The terms $\sum_j P_{\text{UE}, \mathbf{o}_j \rightarrow i}$ represent the powers received by users from all other base stations $j$ whose signals are transmitted on the same PRB allocation at the same time as $i$-th UE in the serving base station and are therefore \textit{inter-cell interference} (ICI).} {\color{black} We treat the ICI as Gaussian noise with power bounded by $( \vert \mathcal{C} \vert - 1 ) P_\text{BS}^\textrm{max} / N_\text{PRB}$ where $P_\text{BS}^\textrm{max}$ is the maximum indoor BS power and $N_\text{PRB}$ is the number of physical resource blocks available in the indoor BS.}

 
\subsection{Problem Formulation}
{\color{black} We formulate the problem {\color{black}addressed by the tuning framework through tuning actions and states} as an optimization problem with a given objective as
\begin{equation}
\label{eq:optimization}
\begin{aligned}
& \underset{\mathbf{a} = [a_1, a_2, \ldots, a_\tau]^\top}{\textrm{minimize:}} \qquad &  \Omega({\bf a},{\color{black} \mathcal{N}; N_\text{UE})}\\
& \text{subject to:} \qquad & c_i \le c, & \qquad i \in \{ 1, \ldots,  N_{\rm UE} \} \\
                && a_t \in\mathcal{A}, & \qquad t \in \{1, 2, \ldots, \tau\} \\
\end{aligned}
\end{equation}
which is not convex due to the non-convexity of the constraints.
We require to find a near-optimal sequence of actions for the problem \eqref{eq:optimization}. 
This sequence of actions {\color{black}$\mathbf{a} = [a_i]_{i=1}^\tau, a_i\in\mathcal{A}$} optimizes a {\color{black}certain network tuning objective $\Omega(\cdot)$ while keeping the tuning effort cost for the $i$-th user $c_i$ bounded} above (or below).  {\color{black}The tuning effort cost is related to a tuning state as we show later.  The set of network events $\mathcal{N}$ are sampled from a random distribution.  This objective $\Omega(\cdot)$ is therefore an expectation}.  To solve this problem, we provide an RL framework and propose RL-based algorithms, thereby avoiding the exhaustive search for all possible {\color{black}network tuning action sequences.  We will further discuss the tuning actions and the states derived from the tuning effort cost in Sections~\ref{sec:reinf} and \ref{sec:algorithms}.}
\section{Reinforcement Learning}\label{sec:reinf}
In this section, we introduce Markov Decision Processes and explain the difference between different policies used in reinforcement learning.

\subsection{Markov Decision Process}
To formulate the problem as a RL problem, we define a Markov Decision Process (MDP) which depends on the current state rather than the previous ones. To apply MDP as part of the problem formulation, we have to define the network states, actions, transition probability, and rewards.  The details are as follows:

\begin{itemize}[leftmargin=*]
    \item \textit{States}: The algorithm is in state $s\in\mathcal{S}$ depending on whether the network performance deteriorated, remained steady, or improved.  A state is \textit{terminal} when the state $s$ is the final state or when the objective has been met.   We define $m\triangleq\vert \mathcal{S}\vert$.
    \item \textit{Actions}: An action $a\in\mathcal{A}$ is one of the valid choices that the algorithm can make to change the state of the network from the current state $s$ to the target state $s^\prime$. We define $n\triangleq\vert\mathcal{A}\vert$.
    \item \textit{Transition Probability}: {\color{black} The transition probability  $p(s^\prime\given s,a)$ is the probability of transitioning to the next state $s^\prime$ given a certain action $a$ and state $s$ at a given time}.  These probabilities are not easily obtained in a realistic cellular network with many UEs and actions.  {\color{black}Furthermore, they may not be well-defined in a model-free reinforcement learning problem, such as $Q$-learning \cite{Sutton}.} 
    \item \textit{Rewards}: The reward $r_{s,s^\prime,a}$ is obtained after the algorithm takes an action $a\in\mathcal{A}$ when it is in state $s\in\mathcal{S}$ and {\color{black}moves to state $s^\prime\in\mathcal{S}$} at discrete time step $t\colon t\in\{0,1,\ldots, \tau\}$.  If the action is accepted by the network and brings the network closer to the objective, the reward is positive.  Otherwise, the reward is negative.  We use very small negative rewards (i.e., $r_\text{min}$) to discourage the agent from taking an action.  Once the algorithm meets the objective, the algorithm obtains a large positive reward. The reward can be defined as
        \begin{equation}
{\color{black} r_{s,s^\prime,a}}\triangleq \begin{cases} 
r_0, & \; \text{if } s^\prime = s_0, \forall {\color{black} (s,a)\in\mathcal{S}\times \mathcal{A}} \\
r_1,  & \;  \text{if } s^\prime = s_1 , \forall {\color{black} (s,a)\in\mathcal{S}\times \mathcal{A}} \\
\vdots \\
r_m, & \;  \text{if } s^\prime = s_m , \forall {\color{black} (s,a)\in\mathcal{S}\times \mathcal{A}}.
   \end{cases}
   \label{eq:rewards}
\end{equation}

\end{itemize}

\begin{figure}[!t]
\begin{adjustwidth}{0cm}{0cm}
\centering
\begin{tikzpicture}[node distance = 5em, auto, thick, scale=2, font=\scriptsize]]
    \node [rectangle, draw, 
    text width=8em, text centered, rounded corners, minimum height=2em,fill=gray!30] (Agent) {Network Tuning Algorithm (Agent)};
    \node [rectangle, draw, 
    text width=8em, text centered, rounded corners, minimum height=1em, below of=Agent] (Environment) {Cellular Network (Environment)};
    
     \path [draw, -latex] (Agent.0) --++ (2em,0em) |- node [text width=5em,near start]{Action\\ $a\in\mathcal{A}$} (Environment.0);
     \path [draw, -latex] (Environment.180) --++ (-2em,0em) |- node [text width=5em, xshift=-0.4cm,yshift=-1.6cm] {Target state\\ $s^\prime\in\mathcal{S}$} (Agent.180);
     \path [draw, -latex] (Environment.north) --++ (0em,0em) -- node [right] {Reward $r_{s,s^\prime,a}$} (Agent.south);
\end{tikzpicture}
\end{adjustwidth}
\caption{The agent-environment interaction in a reinforcement learning framework.}
\label{fig:rl}
\end{figure}
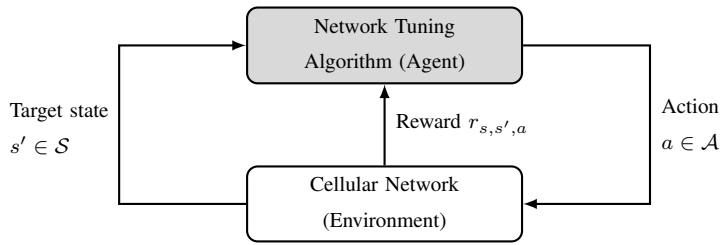

Knowing the reward and the transition probability are key to deriving the optimal decisions.  
Due to the difficulty in estimating the transition probability, we use RL as a solution for this estimation.  The advantage of using an RL-based approach is that this approach can learn from previous experience in a trial-and-error setting, and thus can choose the appropriate actions without an explicit transition probability.

The algorithm behaves as the RL \textit{agent} and interacts with the cellular network elements (i.e., the \textit{environment}) as shown in Fig.~\ref{fig:rl}.  At each time step $t$, the agent plays a certain action $a$ and is in a certain state $s$.  The agent moves to a target state $s^\prime$ and receives a reward $\color{black}r_{s,s^\prime,a}$.  {\color{black} We use the $Q$-learning algorithm of reinforcement learning.} We denote $\color{black}Q_t(s,a)$ as the state-action value function at time step $t$ (i.e., the expected discounted reward when starting in state $s$ and selecting an action $a$).  Our goal is to find a near-optimal solution that maximizes this state-action value function.   Sections~\ref{sec:ql_pc} and \ref{sec:ql_son} explain how we obtain the state-action value function for different $Q$-learning environments.

\subsection{Policy}
{\color{black} $Q$-learning is an off-policy algorithm \cite{Sutton}.  This means that it allows the use of an arbitrary policy during learning.  A policy $\pi(\cdot)$ can be thought of a mapping between the state of the environment and the action to be taken by the agent.  We define our stochastic policy $\pi(a\given s):\mathcal{S}\times\mathcal{A}\rightarrow [0,1]$.  This also means that the policy is a probability distribution of an action $a$ given a state $s$: $\pi(a\given s) \triangleq p(a_t = a\given s_t = s), a\in\mathcal{A}, s\in\mathcal{S}$ for a given time step $t$.  A transition probability can be written down under a policy $\pi$ \cite{Sutton}.
}

An \textit{episode} is a period of time in which an interaction between the agent and the environment takes place. In our case, this period of time is $\tau$ TTIs. During an episode, the agent makes the decision to maximize the effects of actions decided by the agent. {\color{black}We choose a near-greedy action selection rule to represent our policy $\pi$.  This is because with large maximum episode counts, every action will have been sampled many times ensuring a convergence of the state-action value function \cite{Sutton}.   Other selection rules based on sampling or Bayesian statistics require prior knowledge of the distribution of the rewards $\color{black}r_{s,s^\prime,a}$, which may not be easily attainable.  As a result, there are two modes that are applied as follows:}

\begin{itemize}[leftmargin=*]
    \item \textit{Exploration}: to discover an effective action, the agent tries different actions at random.
    \item \textit{Exploitation}: the agent chooses an action that maximizes the state-action value function.
\end{itemize}

Exploitation is suitable for a stable environment where the previous experience is useful while exploration is more appropriate to make a new discovery.  Given that RL is a dynamic iterative learning algorithm, exploration and exploitation are both simultaneously performed through a trade-off strategy known as the $\epsilon$-greedy strategy \cite{Sutton}.  Here, $\epsilon\colon 0 < \epsilon < 1$ is a tuning hyperparameter and allows to adjust the probability between exploration and exploitation, to take the advantages of both exploration and exploitation in an effective manner.  The agent performs exploration with a probability $\epsilon$ and exploitation with probability of $1-\epsilon$.

\begin{table*}[!t]
\setlength\doublerulesep{0.5pt}
\caption{Network Events  $\mathcal{N}$}
\label{table:network_actions}
\vspace*{-0.1in}
\centering
\begin{threeparttable}
\begin{tabular}{ c|l|c||c|l|c} 
\hhline{======}
$\nu^{\rm in}$ & Indoor & Rate & $\nu^{\rm out}$ & Outdoor & Rate\\
\hline 
0 & Cluster is normal. & $p^{\rm in}_0$ & 0 & Cluster is normal. & $p^{\rm out}_{\nu,0}$ \\
1 & Feeder fault alarm (3 dB loss of signal). &$p^{\rm in}_{\nu,1}$ & 1 & Changed antenna azimuth clockwise. & $p^{\rm out}_{\nu, 1}$ \\
2 & Neighboring base station down. &$p^{\rm in}_{\nu,2}$ & 2 & Neighboring base station is down. &$p^{\rm out}_{\nu,2}$\\
3  & VSWR out of range alarm.&$p^{\rm in}_3$ & 3 & Transmit diversity failed. &$p^{\rm out}_{\nu, 3}$\\
4 & Feeder fault alarm cleared.\tnote{\textdagger} &$p^{\rm in}_{\nu,4}$ & 4 & Feeder fault alarm (6 dB loss of signal). & $p^{\rm out}_{\nu,4}$ \\
5 & Neighboring base station up again.\tnote{\textdagger} &$p^{\rm in}_{\nu,5}$ & 5 & Reset antenna azimuth.\tnote{\textdagger} &$p^{\rm out}_{\nu,5}$\\
6  & VSWR back in range.\tnote{\textdagger} &$p^{\rm in}_{\nu,6}$ & 6 & Neighboring base station is up again.\tnote{\textdagger} &$p^{\rm out}_{\nu,6}$\\
&&&7 & Transmit diversity is normal.\tnote{\textdagger} & $p^{\rm in}_{\nu,7}$\\
&&&8 & Feeder fault alarm cleared.\tnote{\textdagger} & $p^{\rm out}_{\nu,8}$ \\
\hhline{======}
\end{tabular}
\begin{tablenotes}\footnotesize
\item[\textdagger] These actions cannot happen if their respective alarm did not happen first.  {\color{black}VSWR is voltage standing wave ratio.}
\end{tablenotes}
\end{threeparttable}
\end{table*}

\section{Improving Network Performance Algorithms}\label{sec:algorithms}

\renewcommand{\thefootnote}{\fnsymbol{footnote}}

In this section, we show our proposed algorithms and quantitatively describe the changes in the SINR as a result of both the network events in Table~\ref{table:network_actions} and the RL-based algorithm corrective actions which we later introduce in Section \ref{sec:results}. In reality, there are a lot more network  events than the events listed in Table~\ref{table:network_actions}.  We only choose a subset of the events that can be modeled. {\color{black} We choose this subset to increase the tractability in analytical derivations as we will see in this section.  As a result, the effect of this subset is more easily reproducible.} 



\subsection{Indoor Problem: VoLTE Power Control}
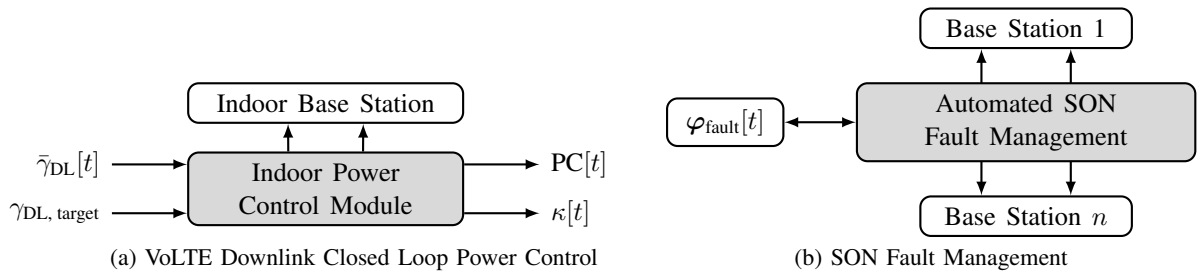
\begin{figure*}[!t]
\begin{adjustwidth}{-0.7cm}{0cm}
\linespread{0.9}
\centering
\subfloat[VoLTE Downlink Closed Loop Power Control]{
\begin{tikzpicture}[style=thick,scale=1]
    \node [rectangle, draw, rounded corners, 
		text width=8em, text centered, minimum height=1em] at (0,0) (bs) {\small Indoor Base Station};
	\node [rectangle, draw, rounded corners, 
		text width=8em, text centered, minimum height=1em, fill=gray!30, below=of bs, yshift=1.5em] (pc) {\small Indoor Power Control Module};

    \path [draw, -latex] (pc.45) -- node {} (bs.331);
    \path [draw, -latex] (pc.135) -- node {} (bs.209);

	\path [draw, latex-] (pc.170) -- ++(-1,0) node[left] {\small $\bar\gamma_\text{DL}[t]$};
	\path [draw, latex-] (pc.190) -- ++(-1,0) node[left] {\small $\gamma_\text{DL, target}$};

	\path [draw, -latex] (pc.10) -- ++(1,0) node[text width=5em] [right] {\small PC$[t]$};
	\path [draw, -latex] (pc.350) -- ++(1,0) node[text width=5em] [right] {\small $\kappa[t]$};
\end{tikzpicture}%

\label{fig:volte}}
\hspace*{-0.4in}
\subfloat[SON Fault Management]{
\begin{tikzpicture}[style=thick,scale=1]

\node [rectangle, draw, rounded corners, 
         text width=6em, text centered, minimum height=0.5em] at (4,1) (bs1) {\small Base Station 1};

\node [rectangle, draw, rounded corners, 
         text width=6em, text centered, minimum height=0.5em,fill=white] at (4,-1.5) (bs2) {\small Base Station $n$};


\path [draw, latex-latex] (bs1.335) -- node {} (bs2.25);
\path [draw, latex-latex] (bs1.205) -- node {} (bs2.155);

\node [rectangle, draw, rounded corners, 
            text width=10em, text centered, below of=bs1,yshift=-0.25cm, fill=gray!30] (fm) {\small Automated SON Fault Management};

\node [rectangle, draw, rounded corners, 
    text width=3em, text centered, minimum height=1em, left of=fm, xshift=-3cm](init) {\small $\bm{\varphi}_\text{fault}[t]$};

\path [draw, latex-latex] (init.0)  -- node [near start] {} (fm.180);

 
\end{tikzpicture}%
\label{fig:reinf_fm}
}
\end{adjustwidth}
\caption{(a) Downlink power control module. $\bar\gamma[t]$ is the {\color{black}effective received} signal to noise plus interference ratio (SINR) at time $t$ fed back to the downlink power control module, which has to maintain the downlink SINR at the receiver at $\gamma_\text{DL, target}$.  
(b) The deep $Q$-learning module interacting with the self-organizing network (SON) and the fault register $\bm{\varphi}_\text{fault}$.  It has to repair the faults in the base station.} 
\label{fig_sim}

\end{figure*}

{\color{black} In this problem, we perform downlink power control without the involvement of the UE sending power commands to the base station.  Rather, the base station autonomously computes the power commands through reinforcement learning.  The base station attempts to power control its transmit power for a single UE at any given TTI $t$ in a round robin fashion.} 
We track the network impairments in this indoor cluster through observing the change in the effective received downlink SINR.  A few events which cause impairments are listed in Table~\ref{table:network_actions}.

Now we can write the \textit{effective received downlink SINR} at a given TTI $t$, $\bar\gamma_{\text{DL}}[t]$ in dB as
\begin{equation}
\label{eq:eff_sinr}
\bar\gamma_{\text{DL}}[t] \triangleq 10\log \left ( \frac{1}{N_\text{UE}} \sum_{i = 1}^{N_\text{UE}} \gamma_{\text{DL}}^{(i)}[t] \right )  \qquad\text{(dB)}.
\end{equation}
{\color{black}The maximization of the individual downlink SINR $\gamma_\text{DL}^{(i)}$ is equivalent to the maximization of \eqref{eq:eff_sinr} because we maximize the SINR of the $i$-th UE (i.e., a single UE) at any given TTI $t$.}


We quantitatively define the improvement of the downlink SINR $\Delta_\gamma$ as the sum of the change in the effective SINR due to the sequence of network events  $\bm{\nu}$ and the sequence of the corresponding algorithmic actions $\mathbf{a}$ over a time period $\tau$ as
\begin{equation}
\color{black}
\label{eq:contribution_2_bf}
\begin{aligned}
    &\Delta_\gamma(\mathbf{a}; \tau, \bm{\nu}) = \bar\gamma_\text{DL}[0] + \sum_{t=1}^\tau \bigg (\delta(a_t\in\mathcal{A}\given \pmb \nu_{1:t}, {\bf a}_{1:t-1}) - \delta(\nu_t\in\mathcal{N}\given \pmb \nu_{1:t-1}, {\bf a}_{1:t-1}) \bigg )
 \end{aligned}
\end{equation}
{\color{black} where $\bar\gamma_\text{DL}[0]$ is the baseline effective received downlink SINR at $t = 0$, ${\bf a} \triangleq [a_1,\dots,a_\tau]^\top$, ${\bf a}_{1:t} \triangleq [a_1,\dots,a_t]^\top$, ${\pmb \nu} \triangleq [\nu_1,\dots,\nu_\tau]^\top$, and ${\pmb \nu}_{1:t} \triangleq [\nu_1,\dots,\nu_t]^\top$.  Also, $\delta(a\given b)$ is the change in the received downlink SINR due to the action $a$ given the network event $b$.} 
Assuming that {\color{black} no prior network event $\bm{\nu}_{1:t-1}$ persists and that all are resolved} in the past by using the proposed algorithm, we can further relax \eqref{eq:contribution_2_bf} to
\begin{equation}
\label{eq:contribution_2}
\begin{aligned}
    &\Delta_\gamma(\mathbf{a}; \tau, \bm{\nu}) = \bar\gamma_\text{DL}[0] + \sum_{t=1}^\tau \bigg (\delta(a_t\in\mathcal{A}\given \nu_t) - \delta(\nu_t\in\mathcal{N}\given \nu_{t-1}) \bigg )
 \end{aligned}
\end{equation}
which makes the change of effective SINR $\Delta_\gamma$ dependent only upon the last state (i.e., Markovian).

We derive the contributions in Table~\ref{table:network_actions} for the indoor network which are only a subset of network faults, as follows:
\begin{itemize}[leftmargin=0.3cm]
\item\textit{Computation of contribution of events} $\nu^\mathrm{in}\in\{1, 3\}$.
When the \textit{voltage standing wave ratio} (VSWR) changes from $v_0$ to $v$ in TTI $t$ due to loose components on the antenna path \cite{6037158}, we compute the change in loss due to return loss as \cite{pozar2000microwave}

\begin{equation}
\Delta L  = 10\log \left (\bigg\vert\frac{v_0 + 1}{v_0 - 1}\bigg\vert \bigg\vert\frac{v - 1}{v + 1}\bigg\vert\right )^2.
\end{equation}
Now we compute the SINR gain (or loss) using (\ref{eq:contribution_2}) as $\delta(\nu^\mathrm{in}_t=3) = -\vert\Delta L\vert$.  Event $\nu^\mathrm{in}_t = 1$ is a special case with $\Delta L = 3\,\text{dB}$.
\item\textit{Computation of contribution of event} $\nu^\mathrm{in} = 2$.
When the neighbor base station $\ell$ is down, we derive the lower bound of the SINR of this event as
\begin{align*}\nonumber
\gamma^{(i)}(\nu_t^\mathrm{in} = 2) &= 
\frac{P_{\text{UE}}^{(i)}}{N_0  + \sum_{j\neq \ell: r_j \in\mathcal{C} \setminus \{ \mathbf{o}_0\}}  {\color{black} P_{\text{UE}, \mathbf{o}_j \rightarrow i}} } \\
& \stackrel{(a)}{\ge} \frac{ P_{\text{UE}}^{(i)}}{N_0  + \vert{\mathcal{C} \setminus  \{ \mathbf{o}_0, \mathbf{o}_\ell\}\vert }  P_{\text{BS}}^{\rm max}  } \\
& \stackrel{(b)}{=}\frac{P_{\text{UE}}^{(i)}}{N_0  + (\vert\mathcal{C}\vert - 2)  P_{\text{BS}}^{\rm max}  }
\end{align*}
where  $P_\text{BS}^{\rm max}$ is the maximum transmit power of the indoor base station. 
$(a)$ comes from that we use the maximum small base station transmit powers instead of the increased received power measured at the UE, and $(b)$ is due to the cardinality of $\mathcal{C}$ being reduced by two: the serving base station $0$ and the neighbor $\ell$ from step $(a)$.
A more relaxed lower bound can be obtained if all the neighboring base stations are down, in this case it is $\gamma^{(i)}(\nu_t^\mathrm{in} = 2) = P_\text{UE}^{(i)} / {\color{black}N_0}$.

{\color{black}
Now, we have
\[
\delta(\nu^\mathrm{in}_t=2) \triangleq  \bar\gamma_\text{DL}[t-{\color{black}N}] - \bar\gamma(\nu^\mathrm{in}_t = 2)
\]%
 where $N > 0$ is the scheduler periodicity, which defines how soon in time would the $i$-th UE voice frames be scheduled again.}

\item\textit{Computation of contribution of events} $\nu^\mathrm{in}\in \{4,5,6\}$.
These events are a result of their respective fault actions being cleared.  Therefore, we reverse the effect of events $1, 2,$ and $3$ respectively.
\end{itemize}

\subsubsection{Proposed RL-based PC} \label{sec:ql_pc}

In an environment with potential wireless signal impairments, such as those shown in Table~\ref{table:network_actions}, PC becomes important to ensure the signal robustness and the usability of the network. We propose a closed loop PC algorithm based on RL.  Closed loop PC can change the transmit signal power to improve the downlink SINR of the $i$-th UE so it meets the target SINR $\gamma_\text{DL, target}$  at any given TTI $t$ one user at a time as in Fig.~\ref{fig:volte}.  For this purpose, closed loop PC sends power commands PC[$t$] to control the signal power over the entire duration of the transmission as follows:
\begin{itemize}
    \item To decrease transmit power {$P_{\rm TX}$} by 1 dB, set $\text{PC}[t] = -1$. 
    \item To keep the transmit power {$P_{\rm TX}$} unchanged, set $\text{PC}[t] = 0$. 
    \item To increase transmit power {$P_{\rm TX}$} by 1 dB, set $\text{PC}[t] = 1$. 
\end{itemize}

{\color{black} This target SINR $\gamma_\text{DL, target}$ can be set through a separate mechanism of power control.  This mechanism can change the target SINR to minimize the received packet error rate  \cite{3gpp25214}.
Let $P_{\text{TX}}^{(i)}$ represent the transmitted power to the $i$-th UE from its serving base station ${\bf o}_0$. 
Modifying the general problem in \eqref{eq:optimization}, we formulate the power control problem as
\begin{equation}
\label{eq:optimization_pc}
\begin{aligned}
&\underset{\mathbf{a} = [a_1, a_2, \ldots, a_\tau]^\top}{\textrm{minimize:}} & \sum_{t =1}^{\tau}\sum_{i=1}^{N_{\rm UE}} P_{\text{TX}}^{(i)}[t]\\
&\text{subject to:} &  \qquad \bar\gamma_{\text{DL}}[t] \ge \gamma_{\text{DL}, \text{target}},\\
                && P_{\text{TX}}^{(i)}[t] \le P_\text{TX}^\textrm{max}, & \qquad t \in \{1, 2, \ldots, \tau\}  \\
                && a_t \in\mathcal{A}
\end{aligned}
\end{equation}
where the transmit power $P_{\text{TX}}^{\color{black}(i)}$ cannot exceed the maximum base station power $P_{\rm TX}^{\rm max}$, and the effective received downlink SINR $\bar\gamma_\text{DL}$ cannot be lower than the target SINR $\gamma_{\text{DL,target}}$.
Accordingly, by solving the problem in \eqref{eq:optimization_pc}, we can minimize the total transmit power during the entire TTI $\tau$ while achieving target SINR for each user.}

Due to closed loop PC, we write $P_\text{TX}$ in dBm at any given TTI $t$ for the $i$-th UE as 
\begin{equation}
 \label{eq:clpceq}
\begin{aligned}
P_\text{TX}^{{(i)}} [t] = \min\! \big (P_\text{BS}^{\rm max},P_\text{TX}^{{(i)}}[t - {N}] + \kappa[t]\text{PC}[t]\big) \qquad\text{(dBm)}
\end{aligned}
\end{equation}
where $\kappa[t]$ is the repetition count (if integer) or step size (if float less than 1) of a power command in a given TTI $t$.  This quantity is decided based on how far the current transmit power is from achieving the target SINR as shown in \eqref{eq:optimization_pc}. PC cannot cause the transmit power to exceed the maximum transmit power of the serving base station.  Furthermore, PC commands can be issued in steps per TTI as governed by $\kappa[t]$.  {\color{black} The actions $a_t$ are mapped to the power control commands as we show later.}

  \begin{algorithm}[!t]
            \small
            \caption{\small VoLTE Downlink Closed Loop Power Control}
            \label{alg:the_pc_alg}
            \DontPrintSemicolon
            \KwIn{Initial downlink SINR value ($\gamma_{\mathrm{DL},0}$) and desired target SINR value ($\gamma_\text{DL, target}$).}
            \KwOut{{\color{black} Near}-optimal sequence of power commands required to achieve the target SINR value during a VoLTE frame, which has a duration of $\tau$, amid network impairments captured.}
            Define the power control (PC) actions $\mathcal{A}$, the set of PC states $\mathcal{S}$, the exploration rate $\epsilon$, the decay rate $d$, and $\epsilon_\text{min}$.\; 
            {\color{black}$\mathbf{Q} := \mathbf{0}_{\vert\mathcal{S}\vert\times\vert\mathcal{A}\vert}$ \tcp*{Zero-initialization}}
            
            $t := 0$  \tcp*{Initialize time}
            $\gamma_\text{DL} := \gamma_\text{DL,0}$ \tcp*{Initialize downlink SINR}
            $(s, a) := (0,0)$ \tcp*{Initialize actions and states}
            \Repeat  {$\gamma$\textsubscript{\rm DL} $\ge$ $\gamma$\textsubscript{\rm DL, target}$\;\mathrm{or}\; t \ge\tau$} {
               $t := t  + 1$\;
               $\epsilon := \max(\epsilon\cdot d, \epsilon_\text{min})$\;
               Sample $r \sim \mathrm{Uniform}(0,1)$\;
              \eIf {$r \le \epsilon$} {
               Select an action $a \in \mathcal{A}$ at random.\;
              } {
               Select an action $a \in \mathcal{A}, a = \arg\max_{a^\prime} \color{black}Q_t(s,a^\prime)$.\;
              }
               Perform action $a$ (power control) on $P_\text{TX}[t]$ and obtain reward $\color{black}r_{s,s^\prime,a}$.\;
               Observe next state $s^\prime$.\;
               Update the table entry $Q_t(s,a)$ as in (\ref{eq:bellman_tabular}). \;
               $s := s^\prime$ \;
            }
            Terminal state reached.  Proceed to the next VoLTE frame.\;
        \end{algorithm}

We model the closed loop PC for VoLTE as a reinforcement learning based algorithm using the standard online (or \textit{tabular}) $Q$-learning as shown in Algorithm~\ref{alg:the_pc_alg}.  The set of actions carried out by the agent is $\mathcal{A} \triangleq \{a_i\}_{i = 0}^{n-1}$ and the set of network states is $\mathcal{S} \triangleq \{s_i\}_{i = 0}^{m-1}$.  Our proposed algorithm attempts to solve the optimization problem (\ref{eq:optimization}).

To derive $Q(s,a)$ {\color{black}at time step $t$}, we build an $m$-by-$n$ table $\mathbf{Q}\in\mathbb{R}^{m\times n}$.  This allows us to use the shorthand notation $Q(s,a) \triangleq [\mathbf{Q}]_{s,a}$ for the state-action value function, which is computed as \cite{Sutton}
\begin{equation}
    \label{eq:bellman_tabular}
    Q_t(s,a)\! = \! (1\!-\!\alpha) Q_{t-1}(s,a)\! +\!  \alpha \!\left [ r_{s,s^\prime,a}\! +\! \gamma \max_{a^\prime} Q_{t-1}(s^\prime,a^\prime) \right ]
\end{equation}
where {\color{black} $Q_t(s,a)$ is the state-action value function at time step $t$,} $\alpha\colon 0 < \alpha < 1$ is the learning rate and determines how aggressive the update of $Q_t(s,a)$ is with respect to $t-1$. Next, $\gamma\colon 0 \le \gamma < 1$ is the discount factor and determines the importance of the predicted future rewards. The reward granted at the current time step is $\color{black}r_{s,s^\prime,a}$. The next state is $s^\prime$ and the next action is $a^\prime$.
{\color{black} For the closed loop PC algorithm, the asymptotic time complexity bound is $\mathcal{O}(mn)$ for $m$ states and $n$ actions \cite{Koenig1993}.  The state-space is exhaustive, since the PC command is either up, down, or unchanged as shown in Table~\ref{table:simulated_volte_actions}.  This make $m$ a fixed quantity.  Therefore, the bound becomes $\mathcal{O}(n)$.}

\subsubsection{Fixed Power Allocation} \label{sec:fixed_pa}

To provide a reference performance, we introduce the \textit{fixed power allocation} (FPA) power control which allows to set the transmit signal power at a specific value.   FPA is our baseline algorithm for performance benchmarking purposes.  It is a common power allocation scheme where the total transmit power is simply divided equally among all the LTE PRBs and is therefore constant
\begin{equation}
P_\text{TX}^{(i)}[t] \triangleq P_\text{BS}^{\rm max} - 10\log N_\text{PRB} + 10\log N_\text{PRB}^{(i)}\qquad\text{(dBm)}  \\
\end{equation}
where $N_\text{PRB}$ is the total number of physical resource blocks in the BS and $N_\text{PRB}^{(i)}$ is the number of available PRBs to the $i$-th UE.


{\color{black}\subsubsection{Maximum SINR}\label{sec:upper_bound}
This is an infeasible greedy algorithm, but is a tight upper bound of performance assuming that we could foresee the future SINRs of an arbitrary UE $i$ ahead of time and that the base}
{\color{black} station power is unbounded above.  In this case, \eqref{eq:optimization} becomes}
\begin{equation}
{\color{black}
\begin{aligned}
    t^* &= \underset{t\in\{1,2,\ldots,\tau\}}{\arg\, \max\;} \Delta_\gamma[t]  \\
    P_\text{TX}^{(i)*} &\stackrel{(c)}{=}  P_\text{TX}^{(i)}[0] + \gamma_\text{DL}^{(i)}[t^*] - \gamma_\text{DL}^{(i)}[0] \\
    &\le    P_\text{TX}^{(i)}[0] + \xi^{(i)}   \\
\end{aligned}
} 
\end{equation}
{\color{black} where $\Delta_\gamma$ only has a parameter $t$, and $\xi^{(i)}\colon \xi^{(i)}\ge 0$ is the foreseen improvement in SINR for the $i$-th UE above its baseline SINR $\gamma^{(i)}_\text{DL}[0]$.  $(c)$ comes from \eqref{eq:clpceq} exploiting that the power gain due to power control commands cannot exceed the difference in target DL SINR.  We show the MDP and the transition probabilities under policy $\pi$ for the indoor problem in Fig.~\ref{fig:mdp_indoor}.
}

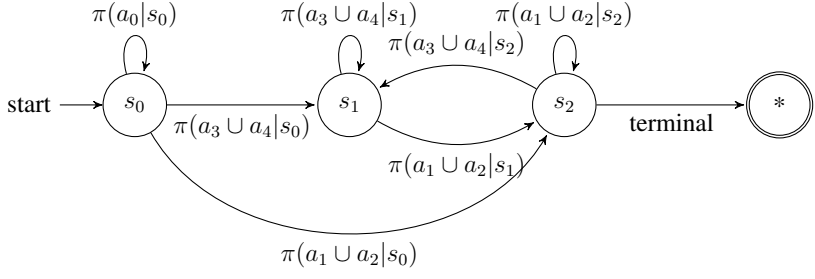
\begin{figure}[!t]
{\color{black}
\centering
    \begin{tikzpicture}[->,>=stealth',scale=1, every node/.style={scale=0.8,node distance=2cm}]
        \node[initial,state] (s0) {$s_0$};
        \node[state, right=of s0] (s1) {$s_1$};	
        \node[state, right=of s1] (s2) {$s_2$};	
        \node[accepting,state, right=of s2] (e0) {*}; 
        \draw[every loop]
            (s0) edge [loop above] node {$\pi(a_0\vert s_0)$} (s0)
            (s1) edge[bend right, auto=right] node {$\pi(a_1\cup a_2 \vert s_1)$} (s2)
            (s1) edge[loop above] node {$\pi(a_3\cup a_4 \vert s_1)$} (s1)
	        (s2) edge[loop above] node {$\pi(a_1\cup a_2 \vert s_2)$} (s2)
	        (s2) edge[bend right, auto=right] node {$\pi(a_3 \cup a_4 \vert s_2)$} (s1)
	        (s2) edge[right, auto=right] node {terminal} (e0)
	        (s0) edge[right, auto=right] node {$\pi(a_3 \cup a_4 \vert s_0)$} (s1)
	        (s0) edge[bend right=60, auto=right] node {$\pi(a_1\cup a_2 \vert s_0)$} (s2)
	    ;
    \end{tikzpicture}
    \caption{Markov decision process and the transitional probabilities under $\pi$ used in the formulation of the power control indoor problem.}
    \label{fig:mdp_indoor}
    } 
\end{figure}

\subsection{Outdoor Problem: SON Fault Management}

In this problem, the algorithm tracks the faults and their impact to the serving base station SINR.  Then we use RL to handle these faults.  We track the faults in the serving base station using a fault register $\bm{\varphi}_\text{fault}\in\mathbb{F}_2^{\vert\mathcal{N}\vert}$, where the $i$-th entry in the register ($i \in \{1, 2, \ldots,\vert\mathcal{N}\vert$\}) corresponds to the network event with identifier $i$ triggered in this cluster as shown in Table~\ref{table:network_actions}.  It is initialized to all logic-0 and set whenever a fault $i$ happens in the network and unset when the fault is cleared.  This algorithm can be implemented at the SON.  Further, we define $\vert\bm{\varphi}_\text{fault}[t]\vert$ as the number of bits that are set to logic-1 in this fault register at TTI $t$. 

We next derive the contributions of the network events in Table~\ref{table:network_actions}.
\begin{itemize}[leftmargin=0.3cm]
\item\textit{Computation of contribution of event} $\nu^\mathrm{out} = 1$.
{\color{black}Changes in antenna azimuth may happen due to interference optimization efforts \cite{5956217} or loose mounting connectors of the antenna at the mounting pole.} When the antenna azimuth changes by an angle $\theta\sim\mathrm{Uniform}(-30^\circ,30^\circ)$, the horizontal plane gain of the antenna changes.  The change is a function of the antenna gain in dB $A(\theta)$ as follows \cite{3gpp36942}
\begin{equation}\nonumber
A(\theta) \!=\! -\min\! \left (\! 12 \left (\frac{\theta}{\theta_{3\text{dB}}}\! \right )^{\!2}\!, A_m \right ), \; -180^\circ\! \le \!\theta\! \le \! 180^\circ
\end{equation}
where $A_m$ is the maximum attenuation of the antenna, $\theta$ is the angle between the direction of interest and the boresight of the antenna, and $\theta_{3\text{dB}}$ is the half-power antenna horizontal beamwidth.  We can now compute the difference in gain as the result of the azimuth change from $\theta_0$ to $\theta$ as $\Delta A(\theta) = A(\theta) - A(\theta_0)$.

\item\textit{Computation of contribution of event} $\nu^\mathrm{out} = 3$.
When the transmit antenna rank $n_t$ decreases, so does the diversity gain and the SINR.

\item\textit{Computation of contribution of event} $\nu^\mathrm{out} \in \{2,4\}$.
(\textit{see analogous computations for the indoor environment contributions}).

\item\textit{Computation of contribution of event} $\nu^\mathrm{out}\in\{5,6,7,8\}$.
These events are a result of their respective fault events being cleared.  Therefore, we reverse the effect of their respective events.
\end{itemize}
\begin{algorithm}[!t]
 \small
  \caption{\small SON Fault Management}
  \label{alg:the_fm_alg}
 \DontPrintSemicolon
  \KwIn{The set of fault handling actions $\mathcal{A}$ in a network $\mathcal{C}$.}
  \KwOut{{\color{black} Near-}optimal fault handling commands given during an LTE-A frame, which has a duration of $\tau$.}
  Define the fault management states $\mathcal{S}$, the exploration rate $\epsilon$, the decay rate $d$, the discount factor $\gamma$, and minimum exploration rate $\epsilon_\text{min}$.\; 
  $t := 0$ \tcp*{Initialize time}
  $(s, a) := (0,0)$ \tcp*{Initialize actions and states}
  $\bm{\varphi}_\text{fault} := [0,0,\ldots,0]$ \tcp*{Initialize fault handling register}
  Randomly initialize $Q$.\;
  Initialize replay memory $\mathcal{D}$.\;
  \Repeat {$\vert\boldsymbol{\varphi}$\textsubscript{\rm fault}$[t]\vert = 0$ {\rm or} $t \ge\tau$} {
  $t := t  + 1$ \tcp*{Next transmit time interval}
  $\epsilon := \max(\epsilon\cdot d, \epsilon_\text{min})$ \tcp*{Decay the exploration rate}
  Sample $r \sim \text{Uniform}(0,1)$\;
  \eIf {$r \le \epsilon$} {
  Select an action $a \in \mathcal{A}$ at random.\;
  } {
  Select an action $a := \arg\max_{a^\prime} Q(s,a^\prime;\bm{\theta}_t)$. \;
  }
  Perform action $a$ to resolve alarm and update $\bm{\varphi}_\text{fault}[t]$.\;
  Obtain reward $r_{s,s^\prime,a}$ from (\ref{eq:rewards}).\;
  Observe next state $s^\prime$.\;
Store experience $e[t] \triangleq (s, a, r_{s,s^\prime,a}, s^\prime)$ in $\mathcal{D}$.\;
Sample from $\mathcal{D}$ for experience $e_j\triangleq (s_j, a_j, r_j, s_{j+1})$.\; 
\eIf {$s_{j+1}$ {\rm is terminal}} {
Set $y_j:= r_j$ 
} {
Set $y_j := r_j + \gamma\max_{a^\prime} Q(s_{j+1}, a^\prime; \bm{\theta}_t)$ 
}
 Perform SGD on $(y_j - Q(s_j, a_j; \bm{\theta}_t))^2$\; 
  $s := s^\prime$\;
}
  Proceed to the next LTE-A frame.\;
 \end{algorithm}

\subsubsection{Proposed RL-based}\label{sec:ql_son}
We propose Algorithm~\ref{alg:the_fm_alg} which is a deep RL-based approach. With a network having $\vert\mathcal{C}\vert$ base stations each having at least $\vert\bm{\varphi}_\text{fault}\vert = F$ faults,  a lower bound of required entires in a tabular $Q$-learning of  $F  \vert\mathcal{C}\vert     \vert\mathcal{S}\vert$  is required.  In networks with thousands of base stations and alarms, the tabular $Q$-learning method to keep track of the state-action values in a table may not scale, hence the use of the \textit{deep $Q$-network} (DQN). 
{\color{black} In fact, with the size of the required tables having millions of elements $(F\vert \mathcal{C} \vert \vert \mathcal{S}\vert)$, the efficiency of tabular $Q$-learning is lower than that of DQNs for two reasons: 1) the latter's ability to learn (i.e., update more weights) faster \cite{lin} and 2) unwanted feedback loops due to correlated sampling of experience may arise and the parameters could get stuck in a poor performing local minimum \cite{mnih2013playing}.  This is in contrast to indoors, where indoor networks typically have a much smaller site count requirement (in magnitude of ones or tens in a building) compared to outdoor networks by design. Therefore, for a matter of convenience, we use tabular $Q$-learning for the indoor problem to provide a lower computational overhead compared to the DQN where computing its weights is burdensome.
}
Fig.~\ref{fig:reinf_fm} shows the interaction of the DQN with the SON.  {\color{black} Using \eqref{eq:optimization}, the algorithm consults the DQN for the alarm $\nu^\mathrm{out}\in\mathcal{N}$ that must be handled using an action $a\in\mathcal{A}$ first to maximize the downlink SINR objective.  This algorithm resides at the SON (or any central location) to ensure coherence across all participating base stations.

We can therefore formulate the SON fault management problem using \eqref{eq:optimization} as}
\begin{equation}
\color{black}
\begin{aligned}
& \underset{\mathbf{a} = [a_1, a_2, \ldots, a_\tau]^\top}{\textrm{minimize:}}\qquad & \vert{\bm\varphi}_\text{fault}[\tau]\vert  \\
& \text{subject to:} &    \vert{\bm\varphi}_\text{fault}\vert \ge 0 \\
                && a_t \in\mathcal{A}, & \qquad t \in \{1, 2, \ldots, \tau\}
\label{eq:optimization_son}
\end{aligned}
\end{equation}
{\color{black} where $\vert{\bm\varphi}_\text{fault}[\tau]\vert$ represents the number of bits set to logic-1 in the fault register at time $\tau$}.  
{\color{black} The objective of this algorithm is to minimize the number of operational faults despite network fault events. Therefore, $\vert{\bm\varphi}_\text{fault}\vert$ depends on the actions $\mathbf{a}$ and the network events $\bm{\nu}$.  
For an agent with a large number of states and actions, or a few states and actions but for a large number of instances, maintaining a table $\mathbf{Q}$ becomes computationally burdensome  as stated earlier, and function estimation with compact parametrization must be used \cite{Sutton}}.  The use of a deep neural network can help estimate the function $Q^*(s,a)$ without having to build the full table \cite{mnih2013playing}.  Fig.~\ref{fig:dnn} shows the structure of the deep neural network used in our algorithms.  We define the estimated $Q^*(s,a)$ as:
\begin{equation}
    \label{eq:bellman_deep}
    Q^*(s,a) \triangleq \mathbb{E}_{s^\prime} \left [ r_{s,s^\prime,a} + \gamma \max_{a^\prime} Q^*(s^\prime,a^\prime) \Given s, a \right ]\!.
\end{equation}

{\color{black} A quick look at \eqref{eq:bellman_deep} shows that the learning rate $\alpha$, present in the tabular version \eqref{eq:bellman_tabular}, is missing here.  The reason is because \eqref{eq:bellman_tabular} uses $\alpha$ to perform the averaging instead of the expectation operator. For deep $Q$-learning, the learning rate is replaced in \eqref{eq:bellman_deep} with the transitional probabilities $p(s^\prime \given s,a)$ for every new state $s^\prime$ as written down from the policy $\pi$.}

If we define the neural network with its weights at time step $t$ as $\bm{\theta}_t\in\mathbbm{R}^{u\times v}$, then (\ref{eq:bellman_deep}) can be approximated using a function approximator $Q(s,a;\bm{\theta}_t)$ such that $Q(s,a;\bm{\theta}_t) \approx Q^*(s,a)$ as $t\to \infty$.  This deep neural network, also known as the DQN, is trained through minimizing a sequence of convex loss functions
\begin{equation}
    \label{eq:loss}
    L_t(\bm{\theta}_t) \triangleq \mathbb{E}_{s,a} \left [(y_t - Q(s,a;\bm{\theta}_t))^2 \right ]
\end{equation}
where $y_t$ is an estimate obtained from the $Q$-network using its weights at time $t-1$ as
\begin{equation}
    \label{eq:yest}
    y_t \triangleq \mathbb{E}_{s^\prime} \left [ r_{s,s^\prime,a} + \gamma \max_{a^\prime} Q(s^\prime,a^\prime;\bm{\theta}_{t-1}) \Given s, a \right ]\!.
\end{equation}
The weights $\bm{\theta}_t$ are updated after every iteration in time $t$ using the \textit{stochastic gradient descent} (SGD) algorithm.  SGD starts with a random initial value of $\bm{\theta}$ and performs an iterative process to update $\bm{\theta}$ as follows
\begin{align}
    \nonumber
    \bm{\theta}_{t+1} \triangleq \bm{\theta}_t - \eta\nabla L_t(\bm{\theta}_t)
\end{align}
where $\eta\colon 0 < \eta \le 1$ is the step size of SGD and $\nabla L_t(\bm{\theta}_t)$ is the gradient of $L_t(\bm{\theta}_t)$ \eqref{eq:loss} with respect to $\bm{\theta}_t$.
We use a method of SGD called adaptive moments \cite{KingmaB14}. We also use the \textit{rectified linear unit} (ReLU) $x\mapsto \max(x,0)$ as the activation function of each node in the DQN.  The deep learning process repeats for all the episodes.

\subsubsection{Random}
To provide a reference for the non-trivial performance lower bound, we introduce a random approach. SON in this approach randomly clears an active alarm by sampling from the fault register $\bm{\varphi}_\text{fault}$.  We choose the discrete uniform random distribution for the clearing of the alarms in the network since the discrete uniform distribution maximizes the discrete entropy \cite{Cover}. A trivial lower bound of the performance is to do no alarm clearing at all.

\subsubsection{First-In First-Out} 
In this approach, the SON takes actions to handle the faults immediately in the next TTI in the order these faults happen.

\begin{figure}[!t]
\centering
{\color{black}
\begin{tikzpicture}[thick,scale=1, every node/.style={scale=1},   cnode/.style={draw=black,fill=#1,minimum width=3mm,circle},
]
 \node at (3,-3.75) {$\vdots$};
  \node at (6,-3.75) {$\vdots$};
    \foreach \x in {1,...,5}
    {   
  
\pgfmathparse{\x <= 3 ? "s_\x" : "\pgfmathresult"}
\pgfmathparse{\x == 4 ? "\vdots" : "\pgfmathresult"}
\pgfmathparse{\x == 5 ? "s_m" : "\pgfmathresult"}
    
        \node[cnode=gray!20,label=180:${\pgfmathresult}$] (x-\x) at (0,{-\x-int(\x/7)+0}) {};
    }

    \foreach \x in {1,...,4}
    {
        \pgfmathparse{\x<4 ? \x : "H"}
        \node[cnode=gray,label=90:$\theta_{\pgfmathresult,1}$] (x2-\x) at (3,{-\x-int(\x/4)}) {};
        \node[cnode=gray,label=90:$\theta_{\pgfmathresult,2}$] (p-\x) at (6,{-\x-int(\x/4)}) {};
        
    }

	\draw[rounded corners=10pt] (-1,-0.5) rectangle ++(2,-5) node[below] at (0, -5.75) {States $\mathcal{S}$};
	\draw[dashed, rounded corners=10pt] (2,0) rectangle ++(5,-5.5) node[below] at (4.5,-5.75) {Hidden layers};
	\draw[rounded corners=10pt] (8,-1.5) rectangle ++(3,-3) node[below] at (9.5,-4.5) {Outputs};

	\foreach \x in {1,...,3}
	{
      	\pgfmathparse{\x== 3 ? "Q^*(s, a_n)" : "Q^*(s, a_\x)"}
      	\pgfmathparse{\x== 2? "\vdots" : "\pgfmathresult"}
     
        	\node[draw=black,circle,label=0:${\pgfmathresult}$] (s-\x) at (9,{-\x-int(\x/4)-1}) {};
	}
	    
    \foreach \x in {1,...,4}
    {   \foreach \y in {1,...,3}
        {   \draw (p-\x) -- (s-\y);
        }
    }
    
    \foreach \x in {1,...,4}
    {   \foreach \y in {1,...,4}
        {  \pgfmathparse{\x<4 ? \x : "H"}   
	        \draw (x2-\x) -- (p-\y) ; 
        }
    }
    \foreach \x in {1,...,5}
    {   \foreach \y in {1,...,4}
        {  \draw (x-\x) -- (x2-\y);
        }
    }
\end{tikzpicture}
} 
\caption{Structure of the neural network used for the Deep $Q$-learning Network implementation with two hidden layers each of dimension $H$.  Here, $(u,v) = (H,2)$, $\vert\mathcal{S}\vert = m, \text{and}\, \vert\mathcal{A}\vert = n$.}
\label{fig:dnn}

\end{figure}
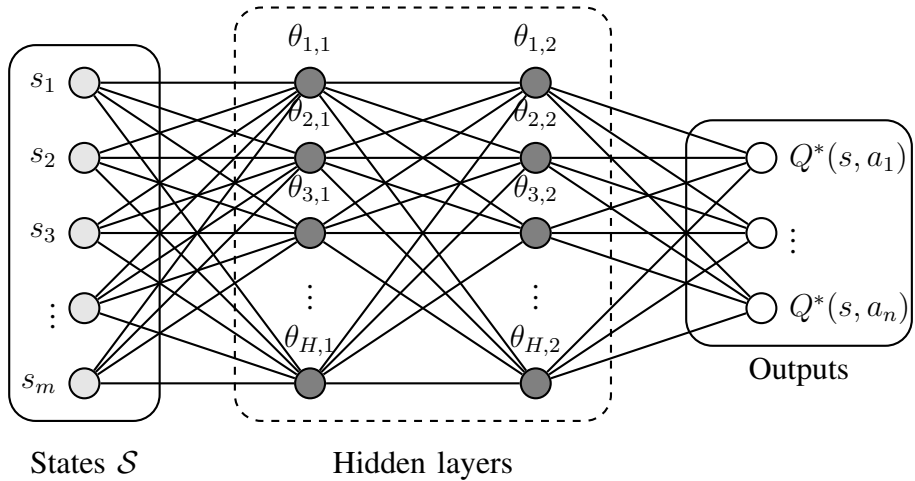


For the random algorithm, an action is randomly sampled from a list of actions; therefore it has a time complexity in $\mathcal{O}(1)$ per iteration or $\mathcal{O}(\tau)$ total.  The First-In First-Out (FIFO) fault-handling algorithm reviews the alarm register every TTI and therefore has a time complexity in $\mathcal{O}(\max(u, \vert\mathcal{C}\vert))$ per time step.  For our proposed algorithm, the time complexity of the DQN backpropagation algorithm is at least in $\mathcal{O}(k(\bm{\theta}) \vert\mathcal{C}\vert\vert\mathcal{A}\vert)$ \cite{scikit-learn}, where $k(\bm{\theta})$ is an increasing function of the depth and number of the hidden layers $\bm{\theta}$.   
Although our proposed algorithm has the highest time complexity cost, the complexity is not dependent on the number of UEs being served, and therefore it is scalable in the number of UEs served in a cluster. {\color{black} We show the MDP and the transition probabilities under the policy $\pi$ for the proposed SON algorithm in Fig.~\ref{fig:mdp_outdoor}.}



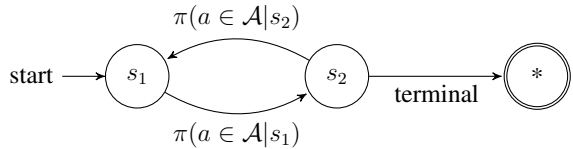
\begin{figure}[!t]
\centering
    \begin{tikzpicture}[->,>=stealth',scale=1, every node/.style={scale=0.8,node distance=1.8cm}]
        \node[initial, state] (s1) {$s_1$};	
        \node[state, right=of s1] (s2) {$s_2$};	
        \node[accepting,state, right=of s2] (e0) {*}; 
        \draw[every loop]
(s1) edge[bend right, auto=right] node {$\pi(a \in \mathcal{A}\vert s_1)$} (s2)
(s2) edge[bend right, auto=right] node {$\pi(a \in \mathcal{A}\vert s_2)$} (s1)
(s2) edge[right, auto=right] node {terminal} (e0)
	    ;
    \end{tikzpicture}
    \caption{\color{black} Markov decision process and the transitional probabilities (under the policy $\pi$ where applicable) used in the formulation of the SON fault handling problem.}
    \label{fig:mdp_outdoor}
\end{figure}


\section{Agent Selection}\label{sec:agent}
{\color{black}
The choice of the implementation of the agent can either be through tabular $Q$-learning or DQN.  In this section, we outline our findings about the choice of one over the other in the RRM problems in network tuning.
\vspace*{-1em}
\subsection{Execution time}

The asymptotic execution time complexity bound of the tabular $Q$-learning algorithm per cell is $\mathcal{O}(mn)$ for $m$ states and $n$ actions \cite{Koenig1993}.  This leads to a total execution time bound in $\mathcal{O}(\vert\mathcal{C}\vert n)$ with the number of states fixed \textit{a priori}.  However, for DQN, the execution time complexity of the DQN backpropagation algorithm is in $\mathcal{O}(k(\bm{\theta}) \vert\mathcal{C}\vert n)$, as discussed earlier.  

Having the number of agent states fixed helps reduce the execution time complexity, as shown in Section~\ref{sec:algorithms}.  Also, we find that the use of tabular $Q$-learning for the indoor problem (or problem with relatively smaller number of cells and users) can provide a lower computational overhead compared to the DQN where computing its weights is burdensome.

\vspace*{-1em}
\subsection{Memory requirement}

The memory requirement in the tabular $Q$-learning is also $\mathcal{O}(\vert\mathcal{C}\vert mn)$.  However, for DQN, the memory requirement is mainly driven by the hidden layers, hence $\mathcal{O}(k(\bm{\theta}))$.
\vspace*{-1em}
\subsection{Stability}
The stability of tabular $Q$-learning necessitates that $\sum_t \alpha_t = \infty $ and $\sum_t\alpha_t^2 < \infty, \forall \alpha$ \cite{580874, jaakkola1994convergence}.  The stability of the DQN is closely related to the stability of the underlying optimizer such as SGD or Adam \cite{KingmaB14}. 

\vspace*{-1em}
\subsection{Convergence}
DQN could get stuck in a poorly performing local minimum or even diverge.  Furthermore, it lacks theoretical convergence guarantees \cite{mnih2013playing}.  However, oscillations and divergence can be avoided using a technique called ``experience replay.''  Experience replay stores experiences in a buffer which are sampled from a uniform distribution.  In tabular $Q$-learning, divergence may occur when updates are not based on trajectories
of the MDP \cite{580874}.  Furthermore, tabular $Q$-learning is prone to initialization bias, where the initial setting of $\mathbf{Q}$ can cause the convergence of the state-action value function to be very slow \cite{5983301}.
\vspace*{-1em}
\subsection{Learning efficiency}

According to \cite{mnih2013playing}, learning directly from consecutive samples is inefficient due to the strong correlation between the samples.  This causes tabular $Q$-learning to be less efficient compared to DQN where the samples are randomized in the experience replay.  This randomization reduces the variance of the updates.  Further, tabular $Q$-learning has a tendency to use the current parameters to determine the next step.  As a result of this, a sequence of updates can cause tabular $Q$-learning to be stuck in a loop.
}

\section{Performance Measures}\label{sec:measures}

\begin{table*}[!t]
\setlength\doublerulesep{0.5pt}
\caption{Reinforcement Learning Hyperparameters}
\label{table:rl_hyperparameters}
\vspace*{-.1in}
\centering
\begin{tabular}{ lr||lr lr } 
\hhline{======}
 \multicolumn{2}{c||}{VoLTE Power Control} &   \multicolumn{4}{c}{SON Fault Management} \\
 \hline
 Parameter & Value & Parameter & Value & Parameter & Value \\
 \hline

One episode duration $\tau$ (ms) & 20 & One episode duration $\tau$ & 10 & Batch size & 32 \\
Discount factor $\gamma$ & 0.995 & Discount factor $\gamma$ & 0.995 & Activation function  & ReLU \\
Exploration rate $\epsilon$ & 1.000& Exploration rate $\epsilon$ & 1.000 & Optimizer & \cite{KingmaB14} \\
Minimum exploration rate $\epsilon_\text{min}$ & 0.010 & Minimum exploration rate $\epsilon_\text{min}$ & 0.010& Hidden layer width $H$ & $24$\\
Exploration rate decay $d$ &  0.99 & Exploration rate decay $d$ &  0.91 & Hidden layer depth & $2$\\
Learning rate $\alpha$ {\color{black} in \eqref{eq:bellman_tabular}} & 0.2 & Optimizer step size $\eta$ & 0.2 \\
Number of states  & 3 & Number of states  & 3 \\
Number of actions   & 5 & Number of actions  & 5  \\
\hhline{======}
\end{tabular}
\end{table*}

\begin{table*}[!t]
\setlength\doublerulesep{0.5pt}
\caption{VoLTE Power Control Algorithm -- Radio Environment Parameters}
\label{table:rf_volte_sim}
\vspace*{-.1in}
\begin{adjustwidth}{0cm}{0cm}
\begin{tabular}{ lrlr } 
\hhline{====}
Parameter & Value & Parameter & Value \\
 \hline
LTE bandwidth & 20 MHz & Base station maximum power  $P_\text{BS}^{\rm max}$ & 33 dBm \\ 
Downlink center frequency & 2600 MHz& Base station initial power setting & 13 dBm \\ 
LTE cyclic prefix  & normal  & Antenna model & omnidirectional \\
Number of physical resource blocks $N_\text{PRB}$ & 100 & Antenna gain $G_\text{TX}$ & 4 dBi \\ 
Cellular geometry & square ($L$ = 10 m) &Antenna height & 10 m\\ 
Propagation model & COST 231 & User equipment (UE) antenna gain & -1 dBi \\ 

Propagation environment & indoor & UE height & 1.5 m \\
Number of transmit antennas & 2  & Max. number of UEs per base station $N_\text{UE}$ & 10 \\
Number of receive antennas & 2  & UE average movement speed & 0 km/h \\
\hhline{====}
\end{tabular}
\end{adjustwidth}
\end{table*}

In this section, we define performance measures to evaluate the proposed algorithms. Different measures are used for the VoLTE power control and SON fault management problems since each problem addresses a different service (i.e., packetized voice vs. high speed data transfer).

\subsection{VoLTE Power Control}

\subsubsection{Voice Retainability}

We define call retainability for the serving cell as a function of the downlink SINR threshold $\gamma_\text{DL, min}$:
{\color{black} 
\begin{equation}
\text{Retainability}  \triangleq 1 - \frac{1}{\tau{\color{black} N_\text{UE}}}\sum_{t = 0}^{\tau}{\color{black}\sum_{i = 1}^{N_\text{UE}}} \mathbbm{1}_{\gamma^{\color{black} (i)}[t] \le \gamma_\text{DL, min}}
\label{eq:retainability}
\end{equation}
where $\gamma^{(i)}[t]$ is the $i$-th UE received SINR obtained at time step $t$.}
\subsubsection{Mean-Opinion Score}
To benchmark the audio quality, we compute \textit{mean-opinion score} (MOS) using an experimental MOS formula \cite{Yamamoto97impactof}.  We obtain the packet error rate from the simulation over $\tau$ frames using the symbol probability of error of a QPSK modulation and a fixed code rate in OFDM \cite{proakis2001digital}.  {\color{black} This enables us to normalize the coding gain of the SINR as a result of this fixed modulation and code scheme.}  We refer to our source code \cite{mycode_volte} for details.

\subsection{SON Fault Management}
\subsubsection{Spectral efficiency}
We evaluate the spectral efficiency with power allocation using the waterfilling algorithm at the transmitter and the zero-forcing equalization at the receiver \cite{VLS-2016}.
The use of spectral efficiency allows us to compare the performance with respect to the upper bound of spectral efficiency of the $M$-QAM modulation used in LTE-A or 5G, since $C \le \log_2 M$.

\subsubsection{Downlink throughput}
We also simulate the average downlink base station throughput and downlink user throughput, which are derived from their cumulative distribution function as follows: peak (95\%), average, and edge (5\%) \cite{link}.

\section{Simulation Results}\label{sec:results}

In this section, we evaluate the performance of the proposed RL-based algorithms via simulations in terms of the performance measures in Section \ref{sec:measures}.
We further explain the intuitions and insights behind these results. The users in the indoor cellular environment follow a homogeneous \textit{Poisson Point Process} (PPP) \cite{bacelli} with intensity $\lambda = 0.5$ users/m\textsuperscript{2}.  The sampled number of connected users is generated using this Poisson distribution while the coordinates of those users are generated using the uniform distribution in a square geometry with length $L$ as in Fig.~\ref{fig:first_case},  which resembles floor plans.  We have four neighboring base stations.  However, for the outdoor environment, we choose a hexagonal geometry as shown in Fig.~\ref{fig:second_case}. {\color{black} We modify the Vienna LTE-A Downlink System Level Simulator 1.9 \cite{VLS-2016} to introduce random faults in the simulated network and invoke fault handling algorithms from a centralized location.  We used the simulator default parameters except for the values in Table~\ref{table:parameters_fm}.  The users in the outdoor network are in an urban environment with both log-normal shadow fading and small-scale fading.}

To find the finite rate as a worst case scenario of predictability \cite{Cover}, we set the occurrence rates of the abnormal network events to be equal and sample from a uniform distribution as
\begin{equation}
\begin{aligned}
    p_{\nu,i} &= p, \qquad \forall i \in\mathcal{N}, i \ge 1 \\
    p_{\nu,0} &= 1 - \sum_{i = 1}^{\vert\mathcal{N}\vert} p_{\nu,i}, \qquad 0 < p_{\nu,j} < 1, \forall j \\
\end{aligned}
\label{eq:rates}
\end{equation}
where $p_0$ denotes the state of normal behavior (i.e, no fault) as shown in Table~\ref{table:network_actions}.  The hyperparameters required to tune the RL-based model are shown in Table~\ref{table:rl_hyperparameters}.  We refer to our source code \cite{mycode_volte,mycode_fm} for further implementation details.

\subsection{VoLTE Power Control}

\begin{table*}[!t]
\setlength\doublerulesep{0.5pt}
\caption{VoLTE Power Control Algorithm -- Simulated Actions $\mathcal{A}$ and States $\mathcal{S}$}
\label{table:simulated_volte_actions}
\vspace*{-.1in}
\centering
\begin{tabular}{ clcl } 
\hhline{====}
Action $a$ & Definition & State $s$ & Definition \\
\hline 
0 & Nothing (this is a transient action). & 0 & No PC issued. \\
1 & Three (PC = $-1$) executed (i.e., $\kappa[t] = 3$).& 1 & PC = $+1$ (Actions $a \in \{ 3, 4\}$ have been played). \\
2 & Single (PC = $-1$) executed (i.e., $\kappa[t] = 1$). & 2 & PC = $-1$ (Actions $a \in \{1, 2\}$ have been played). \\
3 & Single (PC = $+1$) executed.\\
4 & Three (PC = $+1$) executed.\\
\hhline{====}
\end{tabular}
\end{table*}


We run Algorithm~\ref{alg:the_pc_alg} on the indoor cellular network with its parameters in Table~\ref{table:rf_volte_sim}.  
We show the simulated actions and states in Table~\ref{table:simulated_volte_actions}.  The rewards we use in our proposed VoLTE Power Control algorithm are: 
\begin{equation}
r_{s,s^\prime,a}[t;{\color{black}\gamma_\text{DL, target}}] \triangleq \begin{cases} 
r_\text{min}, & \; \bar\gamma_\text{DL}[t] = \gamma_\text{DL, target}  \, \text{not feasible or}\, t\ll\tau \\
	-1, & \; {\text{if}\, s^\prime = s_2\colon} \bar\gamma_\text{DL}[t] < \bar\gamma_\text{DL}[t-N]   \\
	    0, & \;  {\text{if}\, s^\prime = s_0\colon} \bar\gamma_\text{DL}[t] = \bar\gamma_\text{DL}[t-N]   \\
      1, &\;  {\text{if}\, s^\prime = s_1\colon} \bar\gamma_\text{DL}[t] >\bar \gamma_\text{DL}[t-N]  \\
r_\text{max}, & \;  \bar\gamma_\text{DL}[t] = \gamma_\text{DL, target}  \, \text{is met.}
   \end{cases}
   \label{eq:rewards_volte_sim}
\end{equation}
where $N$ is the periodicity of the scheduler.  Based on \eqref{eq:rates}, we set $p^{\rm in}_{\nu, 0} = 5/11$ and $p^{\rm in}_{\nu,1} = p^{\rm in}_{\nu,2} = \ldots = p^{\rm in}_{\nu,6} = 1/11$. We give all faults an equally likely chance of occurrence, which can be considered as the worst case of fault predictability \cite{Cover} and therefore the worst case of the fault handling efficiency.
For the retainability, we choose $\gamma_\text{DL, min} = 0\,\text{dB}$ in (\ref{eq:retainability}). At the SINR of $0$ dB, the calls are likely to drop due to unfavorable channel condition.  We further set $\bar\gamma_{\text{DL},0}$ to $4$ dB and $\bar\gamma_\text{DL, target}$ to $6$ dB.


In the initial episodes with $\epsilon\sim 1$, closed-loop may perform worse than FPA.  However, as $\epsilon\sim\epsilon_\text{min}$, the optimal $Q$\nobreakdash-learning state-action value function (\ref{eq:bellman_tabular}) is learned and the closed loop PC performs better than FPA.  
Fig.~\ref{fig:pc} shows the power command sequence after running the algorithm.  Here, the closed loop PC algorithm causes the base station to change its transmit power consistently (increase, decrease, and unchanged) to meet the desired downlink SINR target as a user is moving in the cell.  On the other hand, FPA has no power commands, which worsens the signal SINR in the presence of signal impairments.

\begin{figure}[!t]
\centering
\includegraphics[scale=0.45]{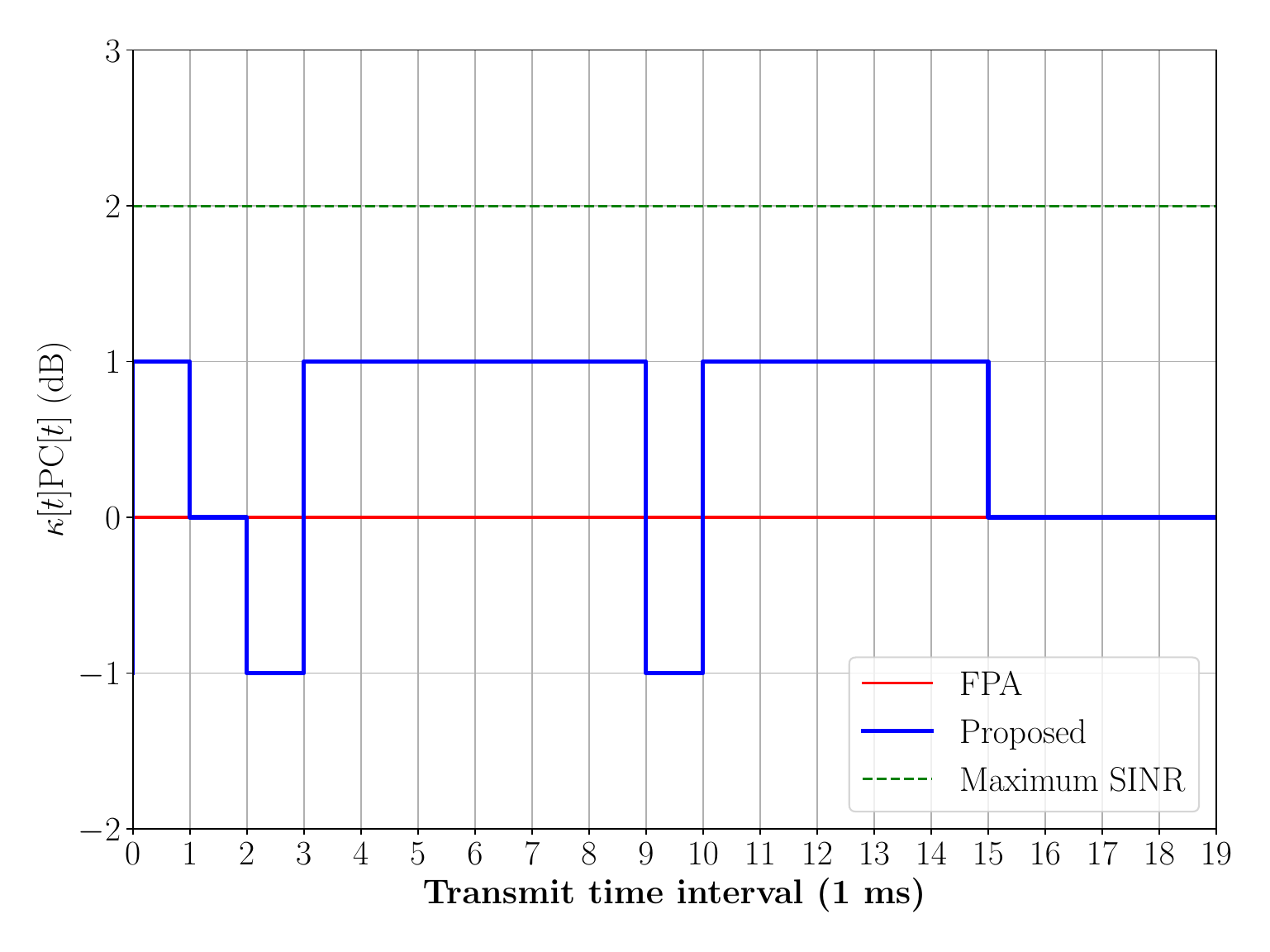}
\vspace*{-1em}
\caption{Power control (PC) sequence.  Unlike fixed power allocation, both the upper bound PC and our proposed closed loop PC using $Q$-learning sends several PCs during the entire VoLTE frame.}
\label{fig:pc}
\end{figure}

\begin{table}[!t]
\setlength\doublerulesep{0.5pt}
\caption{Retainability}
\label{table:retainability}
\vspace*{-.1in}
\centering
\begin{tabular}{ cccc} 
\hhline{====} 
 & Fixed Power Allocation & Proposed & {\color{black}Maximum SINR} \\
\hline
Retainability & 55.00\% & \textbf{78.75\%} & {\color{black}100.00\%} \\ 
\hhline{====}
\end{tabular}
\end{table}


Fig.~\ref{fig:episode826} shows both algorithms where the $Q$-learning based algorithm has learned a near-optimal action-value function.  The closed loop PC pushes the downlink SINR to the target through a near-optimal sequence of power commands.  These power commands are generated from the base station.  The improved retainability and experimental MOS scores due to the closed-loop power control algorithm are shown in Table~\ref{table:retainability} and Fig.~\ref{fig:mos} respectively. 
For the experimental MOS score, we choose a VoLTE data rate of 23.85 kbps and a voice \textit{activity factor} (AF), which is the ratio of voice payload to silence during a voice frame, of 0.7. We refer to our source code \cite{mycode_volte} for further details.

\begin{figure}[!t]
\centering
\subfloat{
\includegraphics[scale=0.21]{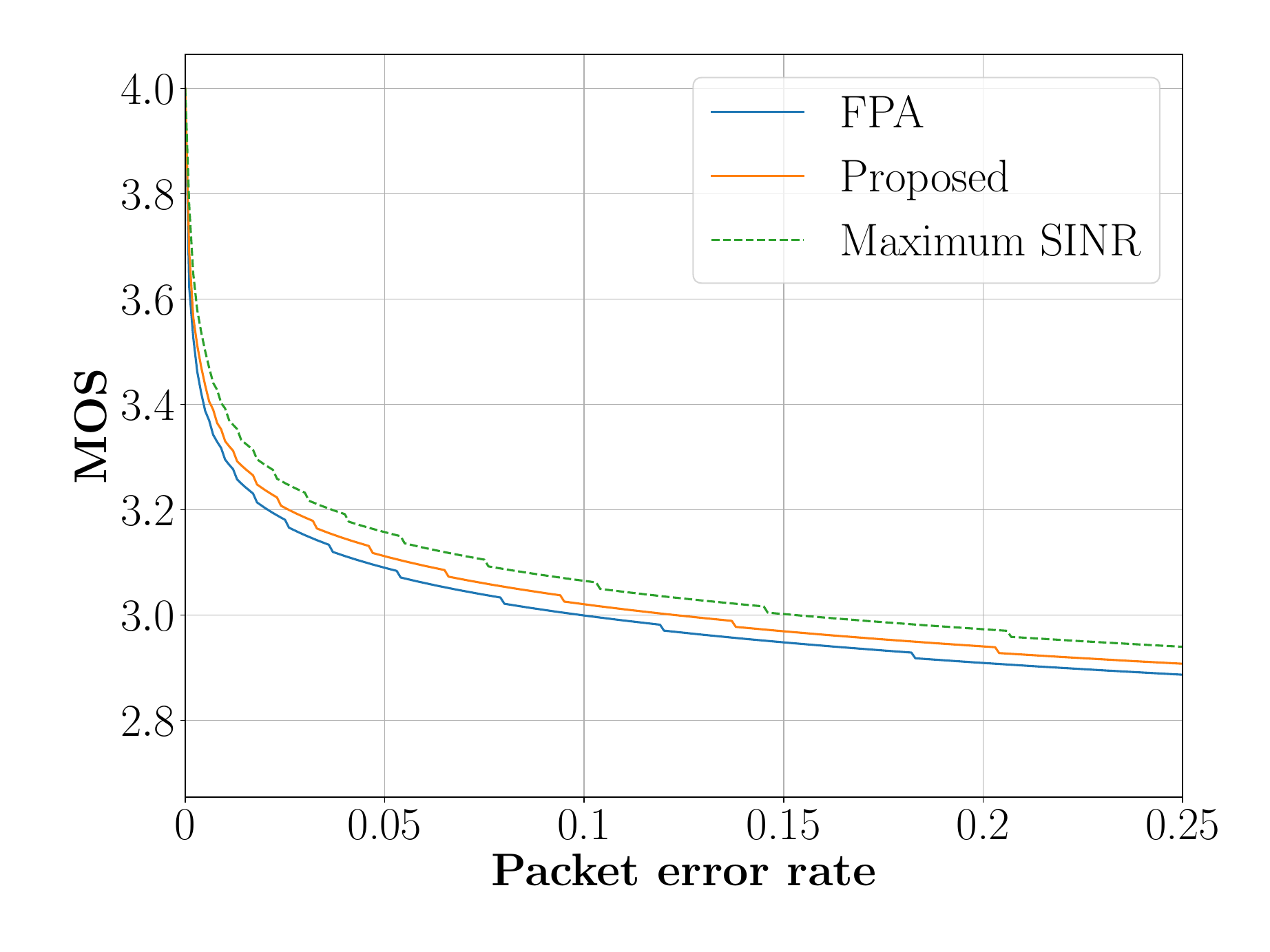}
\label{fig:mos}
}
\hfil
\subfloat{  
\includegraphics[scale=0.25]{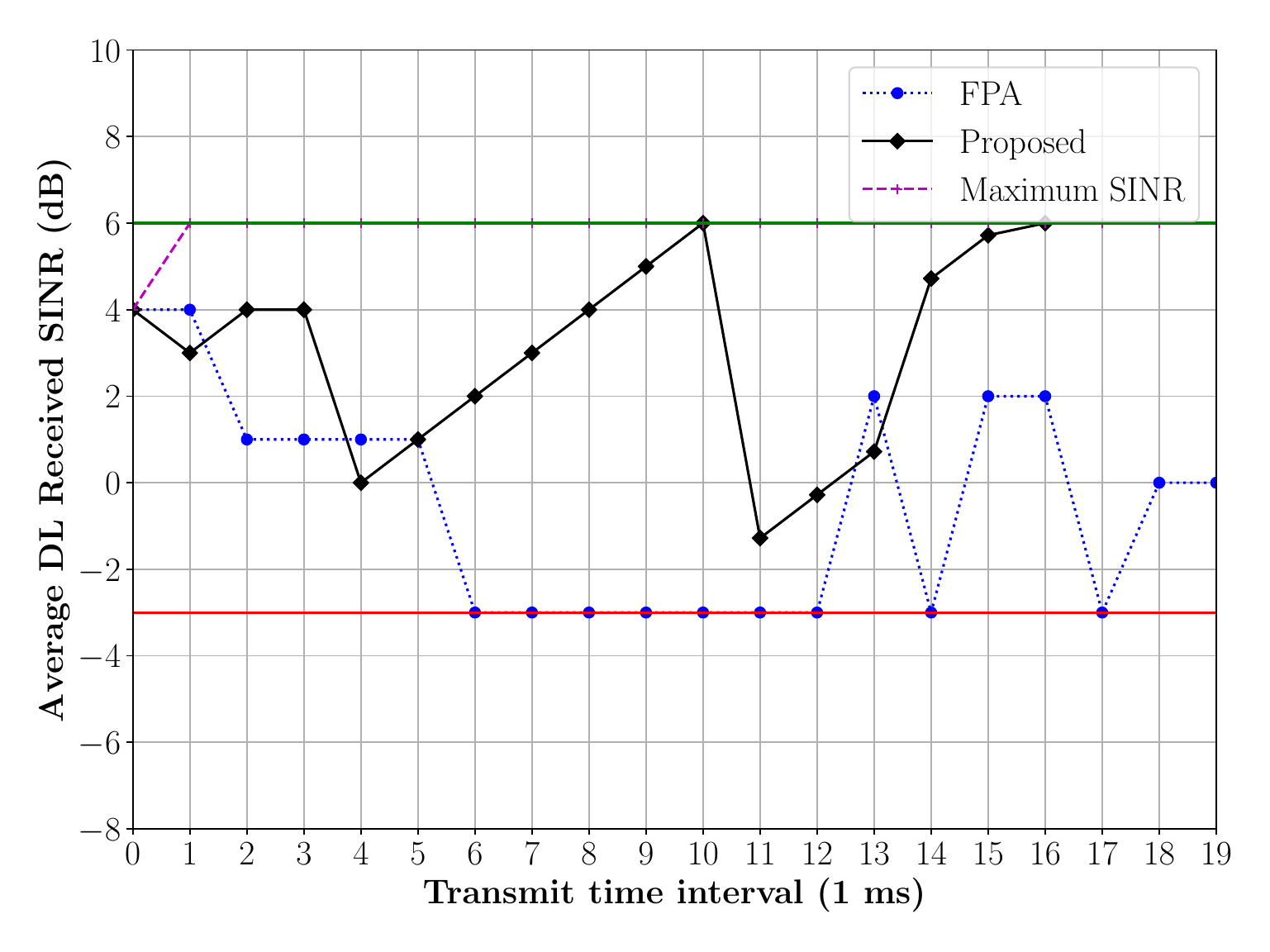}
\label{fig:episode826}
}
\caption{(Left) Mean opinion score (MOS) based on the  voice packet error rate and the experimental formula \cite{Yamamoto97impactof}.  Our proposed closed loop $Q$-learning improved MOS compared to Fixed Power Allocation (FPA). (Right) Downlink SINR improvement vs. simulation time for both our proposed closed loop (CL) power control using $Q$-learning and fixed power allocation (FPA). {\color{black}Here, $\gamma_\text{DL}[0] = 4$ dB and $\xi = 2$ dB. }Green and red lines are  $\gamma_\text{DL, target}$ and  $\gamma_\text{DL, min}$ respectively. CL algorithm reaches the target while FPA does not.}
\label{fig:networks}
\end{figure}



\subsection{SON Fault Management}

\begin{table*}[!t]
\setlength\doublerulesep{0.5pt}
\caption{SON Fault Management Algorithm -- Radio Environment Parameters}
\label{table:parameters_fm}
\vspace*{-0.1in}
\centering
\begin{threeparttable}
\begin{tabular}{ lrlr}
\hhline{====}
Parameter & Value & Parameter & Value\\
 \hline
Bandwidth  & 10 MHz & Downlink center frequency & 2100 MHz \\
LTE cyclic prefix & Normal & Cellular geometry & Hexagonal \\
Inter-site distance &  200m & Scheduling algorithm & Proportional Fair \\
Equalizer & Zero-Forcing & Propagation model & COST231 \\
Propagation environment & Urban & Number of active UEs per BS $q$\tnote{\textdagger}  & $ \{5,10  ,50\}$ \\
BS  antenna model\tnote{\textdagger}  & \cite{3gpp36942} & BS maximum transmit power & 46 dBm \\
BS antenna height & 25 m & BS antenna electrical tilt & 4$^\circ$ \\
Number of BSs in the network & 21 & UE traffic type & Full Buffer \\
MIMO configuration (\# Tx, \# Rx antennas) & $(4,2)$ & Noise power density & -174 dBm/Hz \\
UE average movement speed & 3 km/h   & UE height & 1.5 m \\
Shadow fading margin standard deviation & 8 dB& BS number of sectors per site & 3 \\
\hhline{====}
\end{tabular}
\begin{tablenotes}\footnotesize
\item[\textdagger]  BS is short for \textit{base station} and UE is short for \textit{user equipment}.
\end{tablenotes}
\end{threeparttable}
\end{table*}

We run Algorithm~\ref{alg:the_fm_alg} on the outdoor cellular network with the parameters outlined in Table~\ref{table:parameters_fm}.
We show the simulated actions and states in Table~\ref{table:simulated_son_actions}.  The rewards we use in the SON fault management algorithm are as follows
\begin{equation}
\color{black}
r_{s,s^\prime,a}[t;{\color{black}\boldsymbol{\varphi}_\text{fault}}] \triangleq \begin{cases} 
	-1, & \; {\text{if}\, s^\prime = s_1\colon} \vert\bm{\varphi}_\text{fault}[t]\vert \ge  \vert\bm{\varphi}_\text{fault}[t - 1]\vert\\
      1, &\;  {\text{if}\, s^\prime = s_2\colon} \vert\bm{\varphi}_\text{fault}[t]\vert <  \vert\bm{\varphi}_\text{fault}[t - 1]\vert\\
    r_\text{max}, &\;  \vert\bm{\varphi}_\text{fault}[t]\vert = 0\,\text{(objective is met).}\\
   \end{cases}
   \label{eq:rewards_fm_sim}
\end{equation}

We use a MATLAB-based simulator to generate the LTE network configured in Table~\ref{table:parameters_fm} with reproducibility \cite{VLS-2016}.   We implement Algorithm~\ref{alg:the_fm_alg} using both MATLAB and Python \cite{mycode_fm}.  
In LTE or 5G, the duration of 1 TTI is equal to 1 ms. 
Using \eqref{eq:rates}, we compute the rates in Table~\ref{table:network_actions} as $p^{\rm out}_{\nu,0} = 5/9$ and $p^{\rm out}_{\nu,1} = p^{\rm out}_{\nu,2} = p^{\rm out}_{\nu,3} = p^{\rm out}_{\nu,4} = 1/9$.
For $q \in \{5, 10, 50\}$, we compare the performance of the algorithms in Table~\ref{table:son_performance}.
The random algorithm leads to the worst performance regardless of the number of the UEs per base station $q$ as the order through which the faults are handled is not optimal. Our proposed algorithm outperforms the average downlink spectral efficiency of all algorithms regardless of the number of UEs per base station since the action-value function (\ref{eq:bellman_deep}) has learned an improved fault handling method.


\begin{table*}[!t]
\setlength\doublerulesep{0.5pt}
\caption{SON Fault Management Algorithm -- Simulated Actions $\mathcal{A}$ and States $\mathcal{S}$}
\label{table:simulated_son_actions}
\vspace*{-.1in}
\centering
\begin{tabular}{ clcl } 
\hhline{====}
Action $a$ & Definition & State $s$ & Definition \\
\hline 
0 & No actions issued. & 0 & No actions issued. \\
1 & Faulty neighbor base station is up again. & 1 & Number of active alarms has increased. \\
2 & Serving base station transmit diversity enabled.  & 2 & Number of active alarms has decreased. \\
3 & Serving base station losses recovered.  \\
4 & Serving base station azimuth set to default value.\\
\hhline{====}
\end{tabular}
\end{table*}

We observe that when the base station serves a low number of users, our proposed algorithm outperforms the random algorithm as a lower bound and outperforms the FIFO algorithm. However, as the base station serves more UEs ($q=50$), the performance of all algorithms becomes similar since the cellular resources are near depleted at high base station load (i.e., capacity exhaustion) and therefore clearing alarms does not lead to significant performance improvements. The higher the base station load, the more challenging the SINR improvement is due to the increased inter-cell interference component in (\ref{eq:sinr_final}), and the spectral efficiency tends to have almost no significant variation.


\begin{table*}[!t]
\setlength\doublerulesep{0.5pt}
\caption{Cluster downlink user throughput, average cell throughput, and average spectral efficiency for three different SON Fault Management algorithms: Random, FIFO, and proposed}
\vspace*{-.1in}
\label{table:son_performance}
\centering
\begin{threeparttable}
\begin{tabular}{>{\centering\arraybackslash}p{0.058\textwidth}lccc|ccc|ccc } 
 \cline{3-11}
  & & \multicolumn{3}{c}{Random} & \multicolumn{3}{c}{FIFO} & \multicolumn{3}{c}{Proposed}  \\
  \hline
UEs & Metric & Peak &  Average & Edge  & Peak & Average & Edge  & Peak & Average & Edge \\
\hline
 & UE throughput [Mbps]  & 6.96 & 3.93 & 1.45 & 7.13 & 3.93 & 1.40 & \textbf{7.13} & \textbf{3.93} & \textbf{1.40}  \\
$q = 5$ & Average cell throughput [Mbps] & - & 19.62 & - & - & 19.65 & - & - & \textbf{19.65} & - \\
 & Average SE of UEs [bits/c.u.]\tnote{\textdagger} & - & 2.34 & - & -  & 2.38 & - & - & \textbf{2.38} & - \\

\hline
 & UE throughput [Mbps] & 3.48 & 1.78 & 0.53 & {3.52} & {1.79} & {0.54} & \textbf{3.55} & \textbf{1.84} & \textbf{0.58} \\
$q = 10$ & Average cell throughput [Mbps]  & - & 17.77 & - & - & 17.95 & - & - & \textbf{18.37} & - \\
 & Average SE of UEs [bits/c.u.] & - & 2.21 & - & -  & 2.23 & - & - & \textbf{2.28} & - \\
\hline
 & UE throughput [Mbps]  & 0.68 & 0.38 & 0.13 & 0.68 & 0.38 & 0.13  & 0.68 & 0.38 & 0.13  \\
$q = 50$ & Average cell throughput [Mbps] & - & 18.89 & - & - & 18.90 & - & - & 18.90 & - \\
 & Average SE of UEs [bits/c.u.] & - & 2.36 & - & -  & 2.38 & - & - & {2.38} & - \\
\hhline{===========}
\end{tabular}
\begin{tablenotes}\footnotesize
\item[\textdagger]  SE is short for \textit{spectral efficiency} and UE is short for \textit{user equipment}.
\end{tablenotes}
\end{threeparttable}
\end{table*}

\section{Conclusion}\label{sec:conclusion}
In this paper, we attempted to solve a downlink SINR maximization problem given the worst case distribution of network fault predictability using RL in both indoors and outdoors cellular environments.  We motivated the need for RL in resolving the faults in these realistic cellular environments.
The proposed solution works by allowing RL to learn how to improve tuning objective functions (i.e., downlink SINR and number of active faults) through exploration and exploitation of various corrective actions.  It does so without the UE involvement.
This is beneficial to both indoor and outdoor realistic networks where operational alarms and signal impairments cause degradation to the downlink SINR because it gives the network a chance to recover from the impairments in an efficient way.
The simulations showed that both tabular and DQN RL-based methods, which we proposed in our framework, can improve the QoE-related performance of the cellular network.
Therefore, the proposed RL-based automated cellular network tuning framework is beneficial for improving the performance and maintaining the end-user QoE in a network with impairments and faults.

\bibliographystyle{IEEEtran}
\bibliography{Reinforcement.bib}

\end{document}